\documentclass[11pt,a4paper]{article}

\usepackage{epsfig}
\usepackage{amsmath}
\usepackage{amssymb}
\usepackage{verbatim}
\usepackage{bm}
\usepackage{dsfont}
\usepackage[footnotesize]{caption}
\usepackage{subfigure}
\usepackage{cite}
\usepackage{textcomp}
\usepackage{calc}
\usepackage[margin=0.9in]{geometry}
\usepackage{bbm}
\usepackage{xcolor}
\usepackage[utf8]{inputenc}
\usepackage{lipsum,tabularx}
\usepackage[usestackEOL]{stackengine}
\edef\tmp{\the\baselineskip}
\setstackgap{L}{\tmp}
\usepackage{makecell}
\setcellgapes{4pt}

\usepackage{hyperref}
\hypersetup{colorlinks=true,urlcolor=blue,linkcolor=blue,citecolor=green!60!black}

\newcommand{\centered}[1]{\begin{tabular}{l} #1 \end{tabular}}

\newcommand{\be}{\begin{equation}}
\newcommand{\ee}{\end{equation}}
\newcommand{\bea}{\begin{eqnarray}}
\newcommand{\eea}{\end{eqnarray}}

\begin{document}

\begin{flushright}
CERN-TH-2020-185\\
HIP-2020-28/TH\\
DESY 20-195
\end{flushright}

\vspace*{2cm}

\begin{center}

\bigskip

{\LARGE\bf  Challenges and Opportunities of Gravitational Wave Searches 
at MHz to GHz Frequencies}

\vskip 1.5cm

\renewcommand*{\thefootnote}{\fnsymbol{footnote}}

{
\small N.~Aggarwal$^{a,}$\footnote[1]{Corresponding authors:\\ Nancy Aggarwal (\href{mailto:nancy.aggarwal@northwestern.edu}{nancy.aggarwal@northwestern.edu}),\\ Valerie Domcke (\href{mailto:valerie.domcke@cern.ch}{valerie.domcke@cern.ch}), Francesco Muia (\href{mailto:fm538@damtp.cam.ac.uk}{fm538@damtp.cam.ac.uk}),\\ Fernando Quevedo (\href{mailto:fq201@damtp.cam.ac.uk}{fq201@damtp.cam.ac.uk}), Jessica Steinlechner (\href{mailto:jessica.steinlechner@ligo.org}{jessica.steinlechner@ligo.org}),\\
Sebastian Steinlechner (\href{mailto:s.steinlechner@maastrichtuniversity.nl}{s.steinlechner@maastrichtuniversity.nl})},
O.D.~Aguiar$^b$,
A.~Bauswein$^c$,
G.~Cella$^d$,
S.~Clesse$^e$,
A.M.~Cruise$^f$,
V.~Domcke$^{g,h,i,*}$,
D.G.~Figueroa$^j$,
A.~Geraci$^k$,
M.~Goryachev$^l$,
H.~Grote$^m$,
M.~Hindmarsh$^{n, o}$,
F.~Muia$^{p,i,*}$,
N.~Mukund$^q$,
{D.~Ottaway}$^{r, s}$,
{M.~Peloso}$^{t, u}$,
{F.~Quevedo}$^{p, *}$,
{A.~Ricciardone}$^{t, u}$,
{J.~Steinlechner}$^{v,w,x,*}$,
{S.~Steinlechner}$^{v,w,*}$,
{S.~Sun}$^{y, z}$,
{M.E.~Tobar}$^l$,
{F.~Torrenti}$^\alpha$,
{C.~Unal}$^\beta$,
{G.~White$^\gamma$}}\\[15mm]

\begin{abstract}
\noindent The first direct measurement of gravitational waves by the LIGO and Virgo collaborations has opened up new avenues to explore our Universe. This white paper outlines the challenges and gains expected in gravitational wave searches at frequencies above the LIGO/Virgo band, with a particular focus on \textit{Ultra High-Frequency Gravitational Waves} (UHF-GWs), covering the MHz to GHz range. The absence of known astrophysical sources in this frequency range provides a unique opportunity to discover physics beyond the Standard Model operating both in the early and late Universe, and we highlight some of the most promising gravitational sources. We review several detector concepts which have been proposed to take up this challenge, and compare their expected sensitivity with the signal strength predicted in various models. This report is the summary of the workshop \emph{Challenges and opportunities of high-frequency gravitational wave detection} held at ICTP Trieste, Italy in October 2019, that set up the stage for the recently launched Ultra-High-Frequency Gravitational Wave (UHF-GW) initiative.
\end{abstract}

\newpage
\vspace*{2cm}
{\small\it{
$^{a}${Center for Fundamental Physics, Center for Interdisciplinary Exploration and Research in Astrophysics (CIERA), Department of Physics and Astronomy, Northwestern University, Evanston, Illinois 60208, USA}, \\
$^{b}${Instituto Nacional de Pesquisas Espaciais (INPE), 12227-010 Sao Jose dos Campos, Sao Paulo, Brazil}, \\
$^c${GSI Helmholtzzentrum f\"ur Schwerionenforschung, 64291 Darmstadt, Germany}, \\
$^d${Istituto Nazionale di Fisica Nucleare, Sezione di Pisa, Largo B. Pontecorvo 3, 56127 Pisa}, \\
$^e${Service de Physique Th\'eorique, Universit\'e  Libre de Bruxelles, CP225, \\ boulevard du Triomphe, 1050 Brussels, Belgium}, \\
$^f${School of Physics and Astronomy, University of Birmingham, Edgbaston, Birmingham, UK}, \\
$^g${Theoretical Physics Department, CERN, 1211 Geneva 23, Switzerland}, \\
$^h${Institute of Physics, Laboratory for Particle Physics and Cosmology, EPFL, CH-1015, Lausanne, Switzerland},\\
$^i${Deutsches Electronen Synchrotron (DESY), 22607 Hamburg, Germany} \\
$^j${Instituto de Fisica Corpuscular (IFIC), University of Valencia-CSIC, E-46980, Valencia, Spain}, \\
$^k${Center for Fundamental Physics, Department of Physics and Astronomy,
Northwestern University,
Evanston, IL, USA}, \\
$^l${ARC Centre of Excellence for Engineered Quantum Systems, Department of Physics, University of Western Australia, 35 Stirling Highway, Crawley, WA 6009, Australia}, \\
$^m${Cardiff University, 5 The Parade, CF24 3AA, Cardiff, UK}, \\
$^n${Department of Physics and Helsinki Institute of Physics, PL 64, FI-00014 University of Helsinki, Finland} \\
$^o${Department of Physics and Astronomy, University of Sussex, Brighton BN1 9QH, UK} \\
$^p${DAMTP, Centre for Mathematical Sciences, Wilbeforce Road, Cambridge, CB3 0WA, UK}, \\
$^q${Max-Planck-Institut f{\"u}r Gravitationsphysik (Albert-Einstein-Institut) and Institut f{\"u}r Gravitationsphysik, Leibniz Universit{\"a}t Hannover, Callinstra{\ss}e 38, 30167 Hannover, Germany}, \\
$^r${Department of Physics, School of Physical Sciences and The Institute of Photonics and Advanced Sensing (IPAS), The University of Adelaide, Adelaide, South Australia, Australia} \\
$^s${Australian Research Council Centre of Excellence for Gravitational Wave Discovery (OzGrav)} \\
$^t${Dipartimento di Fisica e Astronomia `Galileo Galilei' Universit\`a di Padova, 35131 Padova, Italy},\\
$^u${INFN, Sezione di Padova, 35131 Padova, Italy}, \\
$^v${Maastricht University, P.O. Box 616, 6200 MD Maastricht, The Netherlands}, \\
$^w${Nikhef, Science Park 105, 1098 XG Amsterdam, The Netherlands}, \\
$^x${SUPA, School of Physics and Astronomy, University of Glasgow, Glasgow, G12 8QQ, Scotland}, \\
$^y${Department of Physics and INFN, Sapienza University of Rome, Rome I-00185, Italy}, \\
$^z${School of Physics, Beijing Institute of Technology, Haidian District, Beijing 100081, People’s Republic of China}, \\
$^\alpha$ {Department of Physics, University of Basel, Klingelbergstr. 82, CH-4056 Basel, Switzerland}, \\
$^\beta${CEICO, Institute of Physics of the Czech Academy of Sciences, Na Slovance 1999/2, 182 21 Prague, Czechia}, \\
$^\gamma${Kavli IPMU (WPI), UTIAS, The University of Tokyo, Kashiwa, Chiba 277-8583, Japan}.}}
\end{center}

\vskip 1cm

\renewcommand*{\thefootnote}{\arabic{footnote}}
\setcounter{footnote}{0}

\newpage
\tableofcontents

\section{Introduction}
\label{sec:intro}
Gravity and electromagnetism are the only two long range interactions in nature, but over the centuries we have explored the Universe only through electromagnetic waves, covering more than 20 orders of magnitude in frequencies, from radio to gamma rays. The discovery of gravitational waves in 2015 has opened a totally new window to observe our Universe~\cite{Abbott:2016blz}.

Judging by what happens with electromagnetic waves, there should be interesting physics to be discovered at every scale of gravitational wave frequencies. Current and planned projects such as pulsar timing arrays, as well as ground- and space-based interferometers will explore gravitational waves in the well-motivated range of frequencies between the nHz and kHz range. However, both from the experimental and the theoretical point of view it is worth to consider the possibility to search for gravitational waves of much higher frequencies, covering regimes such as the MHz and GHz, see for instance~\cite{Cruise:2012zz}.

A strong motivation to explore higher frequencies from the theoretical perspective is that there are no known astrophysical objects which are small and dense enough to emit at frequencies beyond $10 \,$ kHz. Any discovery of gravitational waves at higher frequencies would thus indicate new physics beyond the Standard Model of particle physics, linked e.g.\ to exotic astrophysical objects (such as primordial black holes or boson stars) or to cosmological events in the early Universe such as phase transitions, preheating after inflation, oscillons, cosmic strings, thermal fluctuations after reheating, etc., see~\cite{Caprini:2018mtu} for a recent review.

For early Universe cosmology, gravitational waves may be the only way to observe various events. In particular for the time between the Big Bang and the emission of the cosmic microwave background radiation, electromagnetic waves cannot propagate freely, whereas, due to the weakness of gravity, gravitational waves decouple essentially immediately after being produced and travel undisturbed throughout the Universe forming a stochastic background that could eventually be detected. Even though it may not be easy to unambiguously determine the concrete cosmological source of a gravitational wave signal, its cosmological nature of the spectrum may be identified, similar to what happened with the original discovery of the cosmic microwave background.

In this context, the existence of a stochastic spectrum in the range from kHz to GHz is well-motivated: causality restricts the gravitational wave wavelength to be smaller than the cosmological horizon size at the time of gravitational wave production. This roughly implies a gravitational wave frequency above the frequency range of the existing laser interferometers Virgo~\cite{AdvVirgo,PhysRevLett.123.231108}, LIGO~\cite{aLIGO2015,PhysRevLett.116.131103,PhysRevD.102.062003,PhysRevLett.123.231107} and KAGRA~\cite{Akutsu:2018axf,PhysRevD.88.043007} for any gravitational wave production mechanism that happens at temperatures larger than $10^{10}$ GeV,\footnote{Cosmological events occurring at lower temperatures can also source such high-frequencies gravitational waves if the typical scale of the source is hierarchically smaller than the horizon at that time.}
assuming radiation domination all the way to matter-radiation equality. In particular, GHz frequencies correspond to the horizon size at the highest energies conceivable in particle physics (such as the Grand Unification or string scale) and phenomena like phase transitions and preheating after inflation would naturally produce gravitational waves with frequencies around the GHz range.

Established gravitational wave detector designs are limited to frequencies up to the kHz range. In particular, resonant mass detectors, going back to the original bar design of Joe Weber~\cite{Weber1967}, focused on isolated high-frequencies, often targeting known millisecond-pulsar frequencies. Similarly, the well-established interferometric gravitational wave detectors LIGO, Virgo and KAGRA cover parts of the high-frequency band up to a few kHz. For the purposes of this white paper, we shall therefore use the expression \emph{high-frequency gravitational waves} to refer to frequencies that are above the LIGO detection band, i.e.\ starting from around {10}~{kHz}. In particular, taking inspiration from the electromagnetic spectrum, we denote the MHz to GHz range by \textit{Ultra High-Frequency Gravitational Waves} (UHF-GWs). Several proposals have been made for pushing the high-frequency end of interferometric detectors into this region, however, detectors for the MHz, GHz and THz frequency bands require radically different experimental approaches.

Over the years there have been isolated attempts to search for gravitational waves of very high-frequencies and a few proposals have been put forward. These new concepts have largely been suggested in the form of theoretical papers with no serious discussion of the potential experimental noise sources that might limit their performance, or occasionally, bench tests of early prototypes. The current status of many of these ideas must be regarded as highly preliminary. The published concepts span a wide range of technologies with no real consensus yet as to where to concentrate the community effort. In addition to the selection of suitable technological pathways towards a serious attempt at a detection at high-frequencies, there needs to be an identification of the most realistic sources and thereby the waveforms and spectra for which such detectors should be optimised. This process demands a close collaboration of theorists and experimentalists.

The goal of this report is to summarise and start a dialogue among the specialised community regarding the importance and feasibility to explore searches for high-frequency gravitational waves. We are aware that this may be a long term goal but are convinced that the physics motivation is strong enough to start a systematic study of the different sources of high-frequency gravitational waves and their potential detectability. It is the purpose of this white paper to put together the different ideas both from theory and experiment to explore the importance of searching for high-frequency gravitational waves. The origin of this initiative was a workshop organised at ICTP in October 2019 `Challenges and Opportunities of High-Frequency Gravitational Wave Detection' where members of the theoretical and experimental communities interested on high-frequency gravitational waves got together to explore the motivations and challenges towards this search. This workshop and the present white paper set the stage for the launch of the Ultra-High-Frequency Gravitational Wave (UHF-GW) initiative\footnote{Check out the \href{http://www.ctc.cam.ac.uk/activities/UHF-GW.php}{\underline{website}} of the initiative.}, whose goals include supporting the testing phase of currently existing detector proposals and stimulating the technological developments necessary to come up with new schemes for gravitational wave detectors at high frequencies.

The remainder of this report is organized as follows: Sec.~\ref{sec:Notation} introduces some basic concepts and notation to discuss different types of gravitational wave sources and to relate them to experimental sensitivities. An overview over gravitational wave sources in the late and early Universe is given in Sec.~\ref{sec:th}, followed by a discussion of different detector concepts in Sec.~\ref{sec:exp}. We conclude in Sec.~\ref{sec:conclusion}. For a summary of the various detector concepts and the corresponding sensitivities see Sec.~\ref{sec:SummarySensitivities} and Tab.~\ref{tab:SummarySensitivity}. For a summary of the various sources see Sec.~\ref{sec:Summary}, Fig.s~\ref{fig:hc_stochastic_summary}, \ref{fig:hc_coherent_summary}, App.~\ref{sec:SummaryTable} and Tab.s~\ref{tab:summary-coherent},~\ref{tab:summary-stochastic}.\\

We collect here a few acronyms that will be used throughout the paper: Gravitational Wave (GW), Ultra High-Frequency Gravitational Waves (UHF-GWs), Cosmic Microwave Background (CMB), Black Hole (BH), Innermost Stable Circular Orbit (ISCO), Big Bang Nucleosynthesis (BBN).

\section{Setting up the notation: comparing different GW sources and detectors}
\label{sec:Notation}

Depending on the source/detector, the strength of GWs, detector noise, and signal-to-noise ratio are described using various different metrics~\cite{Maggiore:1900zz}. In general, before using any given metric, it is important to make sure that it is appropriately defined for the scenario under consideration.
In this section we summarize the relevant quantities and notation. We follow the definition in Ref.~\cite{Allen:1999stochastic} for stochastic strain sources, and definitions in Ref.~\cite{Moore:2014sen} for time-dependent strain sources.

\subsection{Gravitational wave sources at high-frequencies}

\begin{enumerate}

\item For stochastic GWs, for example those coming from cosmological sources, a spectral density prescription is most suitable. The most common models assume that they are approximately isotropic, unpolarized, stationary, and have a Gaussian distribution with zero mean. They can thus be fully defined by the second moment~\cite{Allen:1999stochastic}:
\begin{gather}
    \frac{1}{2}\delta^2(\Omega,\Omega')\delta_{AA'} \delta(f-f')\,S_h(f) \equiv \langle \tilde{h}_A(f,\Omega) \tilde{h}_{A'}^*(f',\Omega') \rangle  \,.
    \label{eq:Sh}
\end{gather}

Here $\tilde{h}_A(f,\Omega)$ is the Fourier transform\footnote{Our convention for the Fourier transform (denoted by a tilde) is $\tilde{h}(f) = \int_{-\infty}^\infty dt \, h(t)\, e^{- 2 \pi i f t} $.} of the time-dependent strain in the GW polarization $A$, solid angle $\Omega$, evaluated at a frequency $f$. $S_h(f)$ denotes the one-sided power spectral density.

The energy-density $\rho_\text{GW}$ in GWs per logarithmic frequency interval is represented by $\Omega_\mathrm{GW}$,
\begin{equation}
    \Omega_\mathrm{GW}(f) = \frac{1}{\rho_c} \frac{\partial \rho_{\rm GW}}{\partial \ln{f}} \,,
\end{equation}
conventionally normalized by the critical energy density $\rho_c = 3 H_0^2/(8 \pi G)$ with $G$ denoting Newton's constant and $H_0$ denoting the Hubble parameter today. We will denote the current value of $\Omega_{\rm GW}$ by $\Omega_{\rm GW, 0}$.

The power-spectral density can be directly related to the 00-component of the stress energy tensor, in turn yielding:
\begin{equation}
    \frac{3 H_0^2}{8\pi^2} \Omega_\mathrm{GW}(f)\, f^{-3}\, \delta^2(\Omega,\Omega')  \delta_{AA'} \delta(f-f')= \langle \tilde{h}_A(f,\Omega) \tilde{h}_{A'}^*(f',\Omega') \rangle \,.
\end{equation}

Often, a dimensionless characteristic strain is assigned to the normalized energy density for stochastic GWs (see for instance~\cite{Thrane:2013_sen,Romano:2017Detection})
\begin{subequations}
    \begin{gather}
        h_{c,\rm sto}(f) \equiv \sqrt{f\, S_h(f)} \,,\label{eq:charstrainstochastic}\\
        \Omega_\mathrm{GW} = \frac{4\pi^2}{3H_0^2}f^2 h_{c,\rm sto}^2(f) \,.    \label{eq:OmegaAmplitude}
    \end{gather}
\end{subequations}

\item For inspiral sources, such as BH mergers, a time-dependent strain $h(t)$ can be obtained directly from Einstein's equations. Inspirals have an evolving frequency evolution, so usually the stationary phase approximation is used to obtain an analytical form for the Fourier transform  $\tilde{h}(f)$~\cite{Maggiore:1900zz}. The characteristic strain for such sources with inspiralling frequency can be defined so as to take the frequency evolution into account in the GW strength~\cite{Moore:2014sen},
\begin{equation}
\label{eq:hc}
    h_{c,\rm insp} = 2 f \tilde{h}(f) \,.
\end{equation}
Assuming that $h_0$ is the amplitude of the GW from the inspiral, i.e.\ the amplitude of the periodic function $h(t)$, this results in the characteristic strain:
\begin{equation}
\label{eq:h0}
    h_{c,\rm insp}(f)   = \sqrt{\frac{2f^2}{\dot{f}}} h_0 \,.
\end{equation}

\end{enumerate}

\subsection{Detectors}

Each detector has a different way of searching for GWs, with different antenna patterns, frequency bands, binning, etc. This should be taken into account when defining the appropriate noise and signal-to-noise metrics. For interferometers (such as LIGO) the impact of spatial antenna patterns is of the order of unity. For simplicity, the detector noise floor is usually specified assuming the noise is stationary and Gaussian (even though in reality it is usually neither). Similar to the discussion of stochastic GWs, this noise floor is specified by using a power spectral density,
\begin{equation}
	\frac{1}{2} \delta(f-f') S_n(f) = \langle \tilde{n}(f) \tilde{n}(f')\rangle \,.
\end{equation}
The angular brackets denote an average over multiple realizations of the system, which is obtained repeating the measure of the noise over several well separated time intervals of the same length, see~\cite{Maggiore:1900zz}.\footnote{This assumes that the system is ergodic, hence it is possible to trade an ensemble average with a time average.} In order to measure this noise,  a fast Fourier transform of the detector noise in the absence of the signal is performed. This measured noise is compared to a numerical model, comprising of the sum of all the noises in the detector. An analysis showing each individual noise source (measured or modeled) summing up to the total measured noise is called the \textit{noise budget}.

Unless otherwise specified, if a detector noise is specified in terms of spectral density, it should be treated as $S_n(f)$ (with $\langle\overline{|n(t)|^2}\rangle = \int \mathrm{d}f\, S_n(f)$) if it is in Hz$^{-1}$, or $\sqrt{S_n(f)}$ if it is specified in $1/\sqrt{\text{Hz}}$. For a visual comparison of signal strengths of inspirals and stochastic signals against detector sensitivities, conventionally a dimensionless noise amplitude as been introduced, denoted by $h_{c,n}$~\cite{Moore:2014sen}
\begin{equation}
\label{eq:DimlessNoiseAmplitude}
    h_{c,n} \equiv \sqrt{f S_n(f)} \,.
\end{equation}

Some experiments looking for long-lived sources (e.g. monochromatic sources or stochastic sources) can choose to average the noise over a long time. This gives an additional boost in signal-to-noise ratio, which is sometimes reported as an enhanced sensitivity
\begin{equation}
	S_{n,\mathrm{int}} = \frac{S_{n}}{{N_\mathrm{avg}}} \,.
\end{equation}
If the detector is operating at a frequency $f_\mathrm{center}$ and it is integrating for a time $T_\mathrm{obs}$ then $N_\mathrm{avg} = T_\mathrm{obs}/T_\mathrm{FFT}$ where $T_\mathrm{FFT}$ is the time duration of each spectrum (assuming the segments can be coherently averaged over the duration $T_\mathrm{obs}$).

\subsection{Signal-to-noise ratio}

Understanding whether a signal is detectable using a particular detector requires development of a metric for the signal-to-noise ratio $\rho$.

\begin{itemize}
\item The most efficient signal-to-noise ratio metric for broadband detection of transient sources uses matched-filtering~\cite{Maggiore:1900zz,allen:2012findchirp,Moore:2014sen}:
\begin{equation}
    \rho^2 = \int_0^\infty \mathrm{d}f~  4 \frac{|\tilde{h}(f)|^2}{S_n(f)} = \int d \ln f \, \frac{|h_{c,\rm insp}(f)|^2}{f S_n(f)} \,.
\end{equation}
If the frequency ranges of both the signal and the detector are sufficiently broad, $d \ln f = {\cal O}(1)$, then $h_{c, \rm insp} \sim h_{c,n}$ roughly corresponds to $\rho = {\cal O}(1)$. This explains why $h_{c, \rm insp}$ and $h_{c,n}$ from Eq.~\ref{eq:DimlessNoiseAmplitude} are useful in assessing the reach of a particular broadband instrument looking for an inspiralling source.

\item For a resonant detector with no sensitivity outside a small bandwidth, this signal-to-noise ratio simply collapses to the single frequency band of detection,
\begin{equation}
    \rho^2_\mathrm{res}(f_\mathrm{center},\Delta f) \sim 4 \Delta f ~  \frac{|\tilde{h}(f_\mathrm{center})|^2}{S_n(f_\mathrm{center})} \,,
    \label{eq:SNR_res}
\end{equation}
indicating that a correspondingly larger threshold value of $\tilde h(f)$ is required to yield a detectable signal at fixed $h_{c,n}$.

\item For detecting approximately monochromatic sources, the signal-to-noise ratio similarly collapses to a single frequency. In this case, $\Delta f$ in Eq.~\eqref{eq:SNR_res} is given by the frequency resolution, i.e.\ either the width of the signal or the detector resolution, whatever is the relevant limiting factor. For searches of monochromatic GWs that last over long times, various astrophysical effects like the Earth's motion need to be taken into account.

\item Detecting stochastic sources usually requires utilizing cross-correlation between two or more GW experiments to distinguish the GW background from the experiment's noise (see also Sec.~\ref{sec:crosscorrelation}). Therefore, defining a meaningful signal-to-noise ratio for detection of stochastic sources requires careful consideration of the noise, location, and alignment of each individual experiment. The signal-to-noise ratio can be increased by using more independent experiments, observing for longer times, and optimizing the size of frequency bins. Usually, the strength of the signal will be much less than the detector noise, so cross-correlation can provide a signal-to-noise ratio greater than 1.
For a simple case of $M>1$ colocated detectors, measuring in a frequency band from $f_\mathrm{min}$ to $f_\mathrm{max}$ for a time $T_\mathrm{obs}$, the signal-to-noise ratio can be written as~\cite{Thrane:2013_sen}:
\begin{equation}
    \rho^2_\mathrm{sto} = T_\mathrm{obs} M(M-1) \int_{f_\mathrm{min}}^{f_\mathrm{max}} df\, \frac{S^2_h(f)}{S^2_n(f)} \,\label{eq:SNRsto}.
\end{equation}
We redirect the reader to Refs.~\cite{Allen:1999stochastic,Thrane:2013_sen,Romano:2017Detection} for a full analysis.
\end{itemize}

\subsection{Comparison of signal strength and noise for narrowband detectors}
\label{sec:ComparisonSignalStrength}

Since most high-frequency detectors are narrowband\footnote{We define a detector to be narrowband if its bandwidth is small enough such that the data is analyzed in a single bin, i.e. the noise is assumed to be frequency independent over the bandwidth and infinite outside the bandwidth.}, here we provide some handy expressions to compare signal strength and detector sensitivity for narrowband detectors. The most natural way to express a detector's sensitivity is in power or amplitude spectral density, \(S_n(f)\) or \(\sqrt{S_n(f)}\). On the other hand for signal strengths, the most natural units can depend on the type of source - dimensionless characteristic strain for inspirals, amplitude or power spectral density for stochastic sources, and wave amplitude for long-lived monochromatic sources. In order to compare the signal strength and detector sensitivity, we often strive to convey them in the same units. Here we will provide two ways to achieve this for narrowband detectors. The underlying principle for both methods is to first write down a reasonable signal-to-noise ratio metric, and use that to derive the appropriate comparable quantity. The signal-to-noise ratio for stochastic and long-lived monochromatic sources will be enhanced due to the integration over the observation time, in contrast with signal-to-noise ratio for transient sources, which will depend on just a single observation.

If we are interested in assessing the utility of a given detector to search for GWs from various types of sources, it would be natural to include the integration-time information in the signal depending on the source. On the other hand, if we wish to compare various detectors' suitability to a given source, it is convenient to include the time and bandwidth information to convert the detector sensitivity to the source units. Here we provide ways to do both.

\begin{enumerate}
	\item \textbf{Inspirals:}
		For inspiralling sources passing through the band of a narrowband detector, the signal-to-noise ratio is shown in Eq.~\eqref{eq:SNR_res}, and the Fourier transform \(\tilde{h}(f)\) is related to the characteristic strain using Eq.~\eqref{eq:hc}.
		We desire \(S_\mathrm{h,res,insp}\) or \(h_{c,n,\rm insp}\) such that
		\begin{subequations}
			\begin{align}
				\rho^2_\mathrm{res,insp}(f,\Delta f) \,= \,& \frac{\Delta f}{f^2} \frac{\left|h_{c,\rm insp}(f)\right|^2}{S_n(f)}\\
				\equiv \, & \frac{S_\mathrm{h,res,insp}(f)}{S_n(f)} \\
				\equiv \,& \frac{\left|h_{c,\rm insp}(f)\right|^2}{\left|h_{c,n,\rm insp}(f)\right|^2} \,,
			\end{align}
			\label{eq:convert_narrorband_insp}
		\end{subequations}
which using Eq.~\eqref{eq:SNR_res} gives
		\begin{subequations}
			\begin{align}
				\sqrt{S_\mathrm{h,res,insp}(f)} \sim &\, 2\sqrt{\Delta f} \left|\tilde{h}(f)\right|  
				\,, 
				\\
				\left|h_{c,n,\rm insp}(f)\right| \sim&\, \sqrt{\frac{f^2}{\Delta f}S_n(f)} 
				\,,
			\end{align}
		\end{subequations}
  where \(\Delta f\) represents the frequency range in which GWs are measured by a given detector, i.e., \(f\pm\Delta f/2\).
	\item \textbf{Stochastic sources:}
		Following the same prescription as above, we write the narrowband version of the signal-to-noise ratio in Eq.~\eqref{eq:SNRsto}:
		\begin{equation}
			\rho^2_\mathrm{res,sto} \sim\, T_\mathrm{obs}  \Delta f\, \frac{S_h^2(f)}{S_n^2(f)}\label{eq:SNR_res_sto} \,.
		\end{equation}
		Using Eqs.~\eqref{eq:SNR_res_sto} and\eqref{eq:charstrainstochastic}, we can obtain
		\begin{subequations}
			\begin{align}
				\sqrt{S_\mathrm{h,res,sto}} \sim& \left(T_\mathrm{obs}  \Delta f\right)^{1/4} \sqrt{S_h(f)} 
				= \left( \frac{N_\mathrm{avg}\Delta f}{f}\right)^{1/4}\sqrt{S_h(f)}
				 \,, \\
				\left|h_{c,n,\rm sto}\right| \sim& \left(\frac{1}{T_\mathrm{obs}\Delta f }\right)^{1/4} \sqrt{f S_n(f)} 
\,.\label{eq:charnoisesto}
			\end{align}
		\end{subequations}
	\item \textbf{Monochromatic sources:}

	For long-lived monochromatic sources, the natural unit to specify the signal strength is just the amplitude of the sinusoidal wave, \(h_0\).
	The signal-to-noise ratio buildup is from the long integration times and can be written as
	\begin{equation}
		\rho^2_\mathrm{res,mono} \sim \frac{\left|h_0\right|^2 T_\mathrm{obs} }{S_n(f)} \label{eq:SNR_res_mono}\,.
	\end{equation}
	Using this, noise-equivalent signal \(\sqrt{S_\mathrm{n,res,\rm mono}}\) or signal-equivalent noise \(h_{0,n,\rm mono}\) can be written  \footnote{As is the case in LIGO and VIRGO, a more complicated analysis is needed if the coherence duration of the detector is shorter than $T_\mathrm{obs}$. In that case, fast Fourier transforms taken over the coherence duration are incoherently averaged. This leads to a modified scaling $\rho^2_\mathrm{res,mono}\sim \frac{ |h_0|^2N^{1/2}_\mathrm{avg} T_\mathrm{FFT}}{S_n}$\cite{Astone:2014method}.}:
	\begin{subequations}
		\begin{align}
			\sqrt{S_\mathrm{n,res,mono}} \sim \left|h_0\right|\sqrt{T_\mathrm{obs}} \,,\\
			\left|h_{0,n, \rm mono}\right| = \sqrt{\frac{S_n(f)}{T_\mathrm{obs}}} \,.
			\label{eq:h0mono}
		\end{align}
	\end{subequations}
	Note that for monochromatic sources, there is no need to invent a characteristic strain, the most `characteristic' strain is the amplitude \(h_0\) itself.
\end{enumerate}

%
\section{Sources}
\label{sec:th}
This section reviews various production mechanisms for GW signals in the high-frequency regime, typically in the range $(\text{kHz} - \text{GHz})$, that fall into two broad classes. In Sec.~\ref{sec:lateU} we discuss sources in our cosmological neighbourhood, which emit coherent transient and/or monochromatic GW signals. In Sec.~\ref{sec:earlyU} we turn to sources at cosmological distances which typically lead to a stochastic background of GWs. We emphasize that all proposed sources, with the notable exceptions of the neutron star mergers discussed in Sec.~\ref{sec:NS} (kHz range) and the cosmic gravitational microwave background discussed in Sec.~\ref{sec:CGMB}, require new physics beyond the Standard Model of particle physics to produce an observable GW signal. Thus, while being admittedly somewhat speculative, these proposals provide unique opportunities to shed light on the fundamental laws of nature, even by `only' setting an upper bound on the existence of GWs in the corresponding frequency range.

\subsection{Overview}
\label{sec:Summary}

\begin{figure}[h!]
    \centering
    \includegraphics[width=1.0\textwidth]{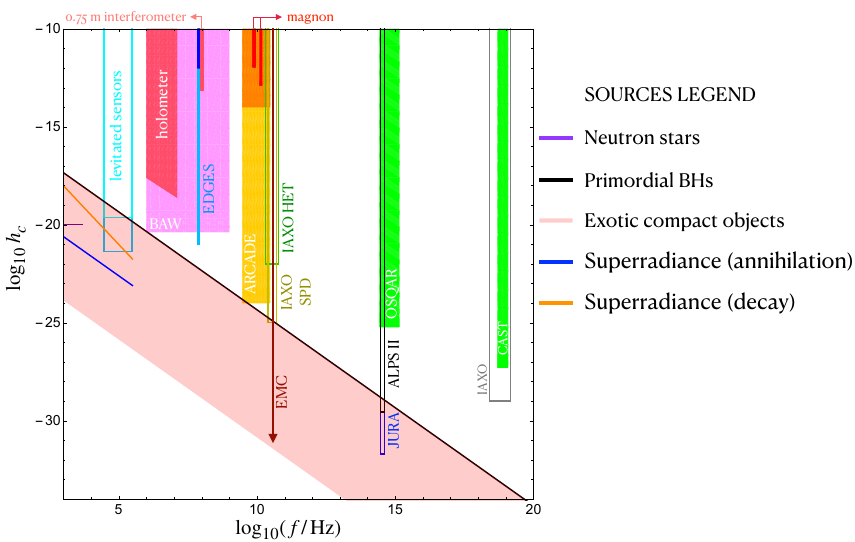}
    \caption{Examples of coherent sources of GWs, see text for details. {Details about the various detector concepts are given in Sec.~\ref{sec:100MHzInterferometers} for the $0.75$ m interferometer and the holometer experiment, Sec.~\ref{sec:OpticallyLevitatedSensors} for the optically levitated sensors, Sec.~\ref{sec:inv_gertsenshtein} for IAXO Single Photon Detector (SPD), IAXO Heterodyne radio receiver (HET), OSQAR, CAST, ALPS II, JURA, EDGES and ARCADE, Sec.~\ref{sec:AmplificationMethods}, Sec.~\ref{sec:BAW} for the Bulk Acoustic Wave Devices (BAW) and Sec.~\ref{sec:GravitonMagnonResonance} the graviton-magnon resonance effect.}}
    \label{fig:hc_coherent_summary}
\end{figure}

\begin{figure}
    \centering
    \includegraphics[width=0.95\textwidth]{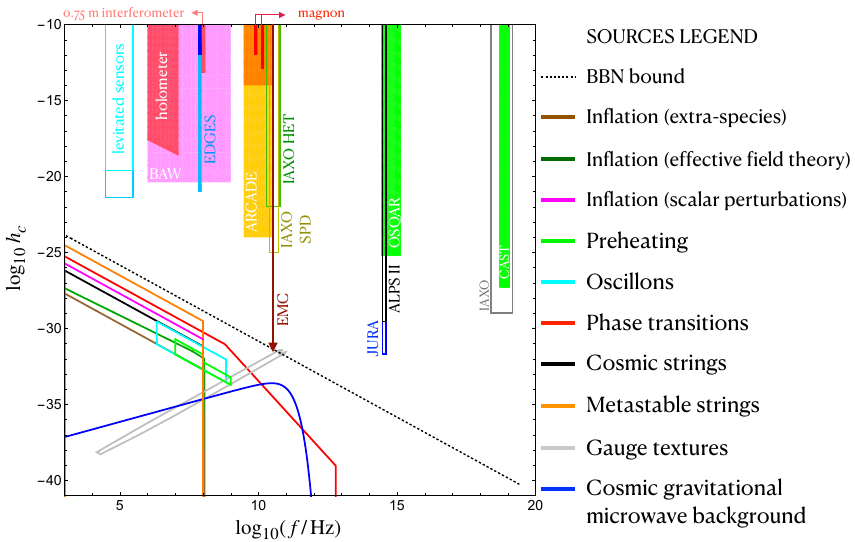}
    \caption{Examples of stochastic sources of GWs, see text for details and the caption of Fig.~\ref{fig:hc_coherent_summary} for the reference to the various detector concept sections.}
    \label{fig:hc_stochastic_summary}
\end{figure}

Fig.~\ref{fig:hc_coherent_summary} and Fig.~\ref{fig:hc_stochastic_summary} summarize a representative selection of the sources which are discussed in more detail in the following subsections. The regions bounded by the colored curves illustrate the region of parameter space which may be covered by the corresponding source for appropriate parameter choices as specified below. Except for the cases of inflation with broken spatial reparametrization symmetry and the cosmic gravitational microwave background they should not be mistaken for GW spectra obtained for a fixed model parameter choice. 

In the same figures, we also indicate the demonstrated (filled boxes) or expected (empty boxes) sensitivity of the detector concepts discussed in Sec.~\ref{sec:detectors}.
In some cases we report two sensitivities for a single detector, using two different intensities of the same color (see for instance the case of levitated sensors), if the sensitivity depends on the details of the future implementation of the detector or on some assumptions needed to place the constraint. In  the case of the levitated sensors the two colors refer to two different versions of the same detector concept: a 1 meter and a 100 meter implementation, see Sec.~\ref{sec:OpticallyLevitatedSensors}, the latter giving a better sensitivity and in the case of the radiotelescopes EDGES and ARCADE, the two sensitivities refer to the weakest and strongest possible cosmic magnetic field, whose value is needed in order to place the constraint, see Sec.~\ref{sec:inv_gertsenshtein}.

The comparison of signal strength and detector sensitivity in these figures should be taken with great caution, and serves as a very rough illustration only. In particular, the signals in Fig.~\ref{fig:hc_coherent_summary} are coherent (and partially transient) signals whereas the signals depicted in Fig.~\ref{fig:hc_stochastic_summary}  are stationary isotropic stochastic signals. A given detector concept will be more or less suitable for these different types of signals, which is not accounted for in this illustration. Further restriction may apply. For example, the quoted sensitivity for the radio telescopes ARCADE and EDGES assumes a cosmological distance between source and observer.
Regarding the possible signals, we have aimed to make realistic estimates of the largest possible signals in different models. This does however not factor in the likelihood of such a signal occurring in the detector lifespan. This is in particular true for the coherent sources, see the respective subsections for details.\\

Fig.~\ref{fig:hc_coherent_summary} shows representative examples of coherent sources. {For simplicity, we take the factor $\sqrt{\frac{2f}{\dot{f}}}$ converting between the amplitude and characteristic strain of a GW to be unity, which is a good approximation at the merging frequency of compact objects. We moreover use a reference value of 10~kpc for the distance to all sources.}
\begin{itemize}
 \item For the ringdown signal of neutron star mergers (Sec.~\ref{sec:NS}) we depict a benchmark at $h_c \simeq 5 \times 10^{-21}$ and $1000<f<5000$ Hz, see Fig.~\ref{fig:spec}.
 \item {For mergers of compact objects, i.e.\ primordial BHs (Sec.~\ref{sec:PBHmergers}) and exotic compact objects (Sec.~\ref{sec:ECOs}) we take the masses of both merging partners to be equal and estimate the maximal signal by determining for each frequency the maximal mass contributing to mergers at this frequency (i.e.\ the mass corresponding to $f = f_\text{ISCO}$ in Eq.~\eqref{eq:fISCO-PBH} or Eq.~\eqref{eq:fISCO-ECO}). For the frequency range depicted, this corresponds to the mass range $(10^{-9}, 1) \, M_\odot$ for primordial BHs. For exotic compact objects, we vary the compactness as $5 \times 10^{-2} < C < 1/2$. The amplitude of the oscillating GW signal is then given by Eqs.~\eqref{eq:h0-PBH} and \eqref{eq:h0-ECO}, respectively. }
 \item {For signals from axion superradiance we consider both the axion annihilation and axion decay channel (see Sec.~\ref{sec:Superradiance}). The frequency of the signal is determined by the axion mass, which is in turn linked to the BH mass by the superradiance condition in Eq.~\eqref{eq:maMBH}. Inserting this into Eq.~\eqref{eq:strain_axion_annihilation} and Eq.~\eqref{GWhg1} and taking $\alpha/l =1/2$, $\epsilon=10^{-3}$ and $M_\text{BH} > M_\odot$ yields the curves depicted.}
\end{itemize}

Fig.~\ref{fig:hc_stochastic_summary} shows models producing stochastic GW signal: interestingly, most of them are concentrated in the UHF band. These are produced in the early Universe and are thus subject to the cosmological constraint on the number of effective degrees of freedom during BBN and at CMB decoupling, see Sec.~\ref{sec:earlyU}.

\begin{itemize}
 \item In certain models, inflation (Sec.~\ref{sec:Inflation}) can yield a signal stretching over a broad frequency range (see Eq.~\eqref{eqfN}), with an amplitude determined by Eq.~\eqref{PGW-frequency} and Eq.~\eqref{nteft}, respectively.
Here in the case of inflation with extra-species we have taken the parameter $\xi$ (defined in Eq.~\eqref{PGW-frequency}) to be bounded by the perturbative limit,
and in the case of inflation described by an effective field theory with broken spatial reparametrization symmetry we have chosen the speed of sound and the spectral tilt to be $c_T = 1$ and $n_T = 0.2$, respectively. Moreover, inflation models with strongly enhanced scalar fluctuations ($P_\zeta \lesssim 10^{-2.5}) $ can source GWs with $\Omega_{\rm GW, 0} \lesssim 10^{-9}$ at second order in cosmological perturbation theory.
\item For preheating (Sec.~\ref{sec:Preheating}), we show typical values for models with parametric resonance in quadratic (right green box) and quartic (left green box) potentials as well as oscillons. In the latter case the frequency is set by the mass of the scalar field through Eq.~\eqref{eq:OscillonFrequency}, where here we have chosen the mass of the scalar field to be $10^{10} \ {\rm GeV} < m < 10^{13} \ {\rm GeV}$ with $X = 100$, while the amplitude is the typical value inferred from numerical simulations.
\item For the cosmic gravitational microwave background, we show the spectrum given by Eq.~\eqref{eq:CGMB} with $T_\text{max} = 10^{16}$~GeV, which is the upper bound on the reheating temperature set by the constraints on the tensor-to-scalar-ratio~\cite{Akrami:2018odb}.
\item For phase transitions (Sec.~\ref{sec:PhaseTransitions}), we assume a fixed latent heat, number of relativistic degrees of freedom and wall velocity. We also assume that sound waves do not last a Hubble time, such that the amplitude scales as the square of the inverse time scale of the transition.
The peak frequency and amplitude are then given by Eqs.~\eqref{eq:PTf} and \eqref{eq:PTOmega}, where we consider transition temperatures $T_* < 10^{16}$~GeV.
\item As an example for topological defects (Sec.~\ref{sec:TopologicalDefects}) cosmic strings lead to a broad spectrum with an amplitude given Eq.~\eqref{eq:plateauStringsNG}, where the string tension for stable cosmic strings is bounded by $G \mu < 10^{-11}$ whereas for metastable cosmic strings it can by as large as $G \mu \simeq 10^{-4}$ above the LIGO frequency range. The spectrum of gauge textures is described by Eq.~\eqref{eq:gaugetextures}, where here we have chosen the symmetry breaking scale to be $10^{12} \ {\rm GeV} < v < 10^{19}~\rm GeV$.
\end{itemize}

\subsection{Late Universe \label{sec:lateU}}

In this section we revise various sources that are relevant for high-frequency GW production and are active in the late Universe. A summary of these sources is given in Fig.~\ref{fig:hc_coherent_summary} and Tab.~\ref{tab:summary-coherent} in App.~\ref{sec:SummaryTable}.

\subsubsection{Neutron star mergers \label{sec:NS}}
\begin{figure}
\begin{center}
    \includegraphics[width=0.6\textwidth]{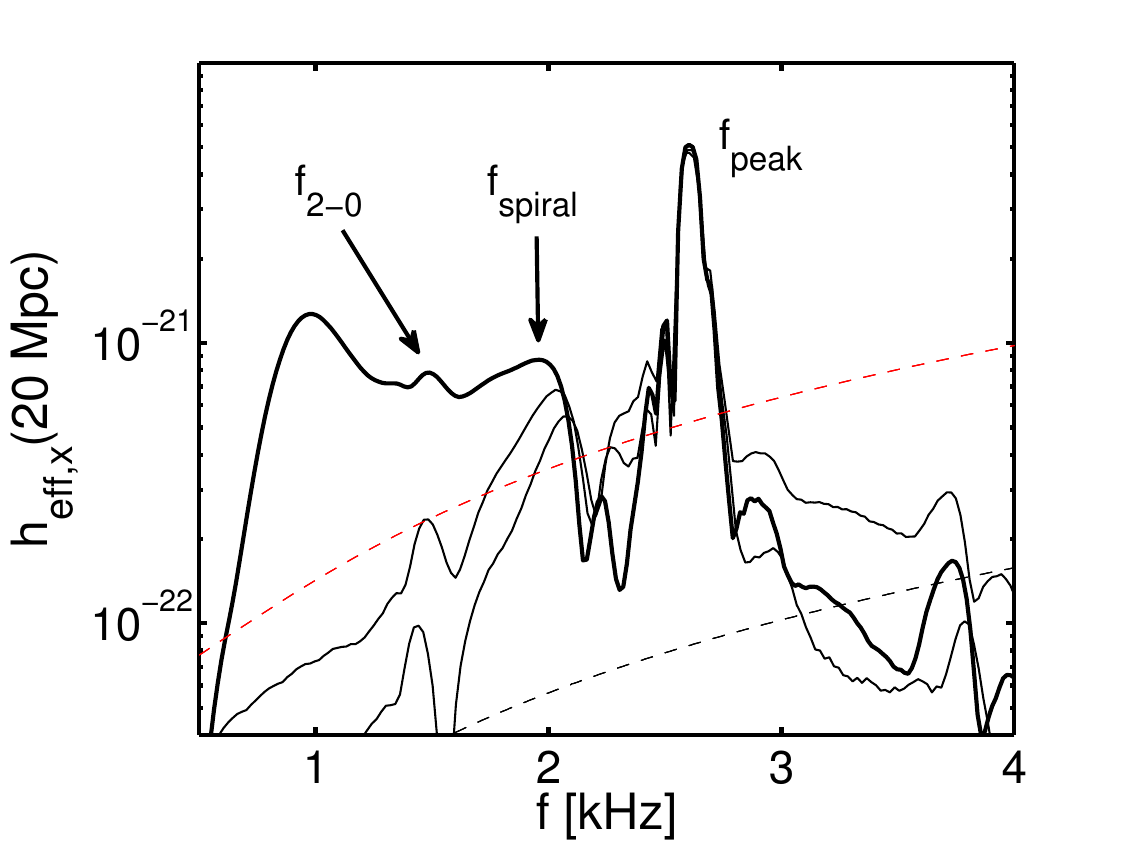}
  \end{center}
  \caption{Typical GW spectrum of the cross polarization of a 1.35-1.35~$M_\odot$ merger along the polar direction at a distance of 20 Mpc. $h_\mathrm{eff,\times} = \tilde{h}_\times(f)\cdot f$ with the Fourier transform of the waveform $h_\mathrm{\times}$ and frequency $f$. $f_\mathrm{peak}$, $f_\mathrm{spiral}$ and $f_{2-0}$ are particular features of the post-merger phase, which can be associated with certain dynamical effects in the remnant. Since the simulation started only a few orbits before merging, i.e. at a relatively high orbital frequency, the power at lower frequencies (below $\sim$ 1 kHz) is massively under-represented in the shown spectrum, and the low-frequency part of the spectrum does not show the theoretically expected power-law decay. The thin solid lines display the spectra of the GW signal of the post-merger phase only revealing that the peaks are indeed generated in the post-merger phase. Dashed lines show the expected unity signal-to-noise sensitivity curves of Advanced LIGO (red) and of the Einstein Telescope (black). Figure taken from~\cite{Bauswein2016}.}
  \label{fig:spec}
\end{figure}

For not too high binary masses the merger of two neutron stars avoids the prompt collapse to a BH and leads to the formation of a massive rapidly rotating and oscillating neutron star remnant. The oscillations of this remnant are very characteristic of the incompletely known equation of state of high-density matter and generate GW emission in the kHz range (see Fig.~\ref{fig:spec}). For instance, the dominant oscillation frequency of the post-merger phase ($f_\mathrm{peak}$ in Fig.~\ref{fig:spec}) scales tightly with the radii of non-rotating neutron stars~\cite{Bauswein2016}. These radii are uniquely determined by the equation of state of neutron stars, and are therefore particularly valuable messengers of the underlying high-density matter physics (see e.g.~\cite{Oertel2017} for a review). Simulation results show a tight correlation between the dominant GW frequency and neutron star radii; a fit to the data for fixed binary masses describes the relation with a maximum residual of only a few hundred meters, allowing for accurate radius measurements.

Subdominant features in the GW spectrum (see Fig.~\ref{fig:spec}) contain additional information about the equation of state and may also reveal the dynamics of the remnant, which is indispensable for a complete multi-messenger interpretation of neutron star mergers.
The presence or absence of post-merger GW emission from a neutron star remnant on its own informs about the outcome of the merger (neutron star or BH). In combination with the measured binary masses, this information allows to constrain the threshold binary mass for prompt BH collapse, which is somewhere in the range $(2.9 - 3.8) \, M_\odot$, depending on the equation of state. This threshold depends sensitively on the maximum mass $M_\mathrm{max}$ of non-rotating neutron stars. Obtaining the threshold mass for prompt BH formation through post-merger GW emission will yield a robust determination of the unknown maximum mass $M_\mathrm{max}$ of non-rotating neutron stars, which is another important equation of state property that probes the very high-density regime~\cite{Bauswein2013}. A robust measurement of $M_\mathrm{max}$ is also relevant for stellar astrophysics since it, for instance, affects the outcome of core-collapse supernovae. Pulsar observations only yield accurate lower bounds on $M_\mathrm{max}$.

Generally, equation of state inference from the post-merger stage is complementary to other constraints, e.g. from the inspiral phase. The complementarity concerns the probed density regime, which is generally higher in the post-merger phase, and methodological aspects. Hence, the detection of post-merger GW emission is of highest importance to understand properties of high-density matter including the opportunity to probe the presence of a phase transition to deconfined quark matter~\cite{Most2019,Bauswein2019}.

The different features of the post-merger GW emission have frequencies in the range $(1 - 5)$ kHz, with the dominant peak between 2 and 4 kHz (see Fig.~\ref{fig:spec}). Simulated injections show that at a distance of 40 Mpc (comparable to that of GW170817) a strain sensitivity of roughly {$\sqrt{S_n} \simeq 3 \times 10^{-24} \, \text{Hz}^{-1/2}$} is required for a detection of the main features~\cite{Torres-Rivas2019}. Hence, measurements can be anticipated with a small sensitivity improvement either of Advanced LIGO/Virgo/KAGRA or with a dedicated high-frequency instrument like NEMO~\cite{ozhf} (see Sec.~\ref{sec:ozgrav}).

\subsubsection{Mergers of light primordial black holes}
\label{sec:PBHmergers}

The low effective spins and progenitor masses of the BH mergers detected by LIGO/Virgo have revived the interest for primordial BHs in the range $(1-100) \, M_\odot$~\cite{Bird:2016dcv,Clesse:2016vqa,Sasaki:2016jop}, which could constitute (part of) the observed dark matter abundance. In this context, detecting a sub-solar mass BH would almost clearly point to a primordial origin.\footnote{See however~\cite{Kouvaris:2018wnh} for another sub-solar BH formation channel, in a specific dark matter scenario.}  The frequency associated to the ISCO when the inspiral GW emission is close to maximal\footnote{The ISCO characterizes the end of the inspiral phase and the beginning of the merger. The merger frequency will be higher, but we use ISCO to demonstrate the ballpark estimated strain analytically.}, is given by
\be
f_{\rm ISCO} = 4400 \, {\rm Hz} \, \frac{M_\odot}{m_1 + m_2} \,,
\label{eq:fISCO-PBH}
\ee
with $m_1$ and $m_2$ the masses of the two binary components and $M_\odot$ denoting the solar mass. A good estimation of the GW strain produced at a given frequency $f$ is provided by the Post-Newtonian approximation~\cite{Antelis:2018sfj}
\be \label{eq:strainPBHs}
h_0 \approx \frac{2}{D} \left( \frac{G \mathcal M }{c^2}  \right)^{5/3} \left( \frac{ \pi f}{c} \right)^{2/3}  \,,
\ee
where $\mathcal M \equiv (m_1 \times m_2)^{3/5}/(m_1+m_2)^{1/5} $ is the binary chirp mass,  $D$ is the distance to the observer, $G$ is Newton's constant and $c$ is the speed of light. For an equal-mass binary and an experiment of strain sensitivity $h_{\rm det}$,  the corresponding astrophysical reach $D_{\rm max}$ is given by
\be
D_{\rm max} \approx 1.6  \, \frac{(m_{\rm PBH}/M_\odot) }{h_{\rm det} \times 10^{20}} {\rm Mpc} \,.
\label{eq:h0-PBH}
\ee
High-frequency GW detectors could therefore detect or set new limits on the abundance of light, sub-solar mass primordial BHs, in particular if they have an extended mass distribution. Frequencies in the range $(10^4 - 10^{12})$ Hz correspond to a primordial BH mass range $(10^{-9} - 10^{-1}) \, M_\odot$. In particular, the planetary-mass range, in which recent detections of star and quasar microlensing events~\cite{Niikura:2019kqi,Hawkins:2020zie,2019ApJ...885...77B} suggest a dark matter fraction made of primordial BHs of $f_{\rm PBH} \sim 0.01$, could be probed in a novel way.

There are two possible formation channels of primordial BH binaries, introduced hereafter:
\begin{enumerate}
    \item \textit{Primordial binaries}:  they come from two primordial BHs that were formed sufficiently close to each other for their dynamics to decouple from Universe expansion before the time of matter-radiation equality~\cite{Nakamura:1997sm,Sasaki:2016jop}. The gravitational influence of one or several primordial BHs nearby prevent the two BHs to merge directly, leading to the formation of a binary.  In some cases, the binary is sufficiently stable and it takes a time of the order of the age of the Universe for the two BHs to merge. If the primordial BHs have a mass spectrum \(\rho(m)\) and are randomly distributed spatially, and that early forming primordial BH clusters do not impact the lifetime of those primordial binaries (a criterion satisfied for $f_{\rm PBH} \lesssim 0.1 $)~\cite{Raidal:2018bbj}, then the today merging rate is approximately given by~\cite{Kocsis:2017yty,Raidal:2018bbj,Gow:2019pok}
    \begin{equation}
        \frac{d\tau}{d(\ln m_1) \,d(\ln m_2)} \approx \frac{1.6 \times 10^6}{\rm Gpc^3 \, yr} f_{\rm PBH}^2 \left(\frac{m_1 + m_2}{M_\odot}\right)^{-\frac{32}{37}} \left[\frac{m_1 m_2}{(m_1+m_2)^2}\right]^{-\frac{34}{37}} \rho(m_1) \rho(m_2)\,,
    \end{equation}
    where $f_{\rm PBH}$ is the integrated dark matter fraction made of primordial BHs, $m_1$ and $m_2$ are the masses of the two binary BHs components and $\rho(m)$ is the density distribution of primordial BHs normalized to one ($\int \rho(m) d \ln m = 1$).\footnote{However, note that if primordial BHs constitute a substantial fraction of the dark matter, then  N-body simulations have shown that early-forming clusters somehow suppress this rate, eventually down to the rates inferred by LIGO/Virgo for $f_{\rm PBH} \simeq 1$.} Assuming $f_{\rm PBH} \times \rho(m) \simeq 0.01$ at planetary masses in order to pass the microlensing limits, and considering only almost equal-mass mergers ($m_1 \sim m_2 \sim m_{\rm PBH}$) that produce the highest strain, one obtains a merging rate of
    \be
    \tau(m_{\rm PBH}) \approx 300  \left(\frac{m_{\rm PBH}}{M_\odot}\right)^{-0.86} {\rm yr^{-1} Gpc^{-3}}\,.
    \ee
    In turn, using Eq.~\eqref{eq:strainPBHs}, one obtains the required GW strain sensitivity to detect one of these merger events per year,
    \be
    h_{\rm max}  \approx 1.7 \times 10^{-22}  \left(\frac{m_{\rm PBH}}{M_\odot}\right)^{0.7} \approx 4.2 \times 10^{-20} \left( \frac{\rm Hz}{f} \right)^{0.7} \,,
    \ee
    which can be typically targeted by GW experiments operating at frequencies from kHz up to GHz, see e.g.~\cite{Wang:2019kaf}.
    \item \textit{Capture in primordial BH haloes}:  the second binary formation channel is through dynamical capture in dense primordial BH halos. As any other dark matter candidate, primordial BHs are expected to form halos during the cosmic history.  The clustering properties typically determine the overall merging rate. For instance, for a monochromatic mass spectrum and a standard Press-Schechter halo mass function, one gets a rate~\cite{Bird:2016dcv}
    \be \label{eq:capturerate}
    \tau \sim f_{\rm PBH} \times \mathcal O(10-100) \, {\rm yr^{-1} \, Gpc}^{-3} \,,
    \ee
    that is independent of the primordial BH mass. However, for realistic extended mass functions, the abundance, size and evolution of primordial BH clusters is impacted by several effects: poissonian noise, seeds from heavy primordial BHs, primordial power spectrum enhancement, dynamical heating, etc. Those can  either boost or suppress the merging rate and make it a rather complex and model-dependent process, subject to large uncertainties (see e.g.~ \cite{Clesse:2016vqa,Ali-Haimoud:2017rtz,Young:2019gfc,Bringmann:2018mxj,Trashorras:2020mwn} for recent studies of primordial BH clustering).  As an alternative of using uncertain theoretical predictions, one can instead infer from LIGO/Virgo observations an upper bound on the primordial BH merging rate of $\tau \approx 1.2 \times 10^4 {\rm yr^{-1} Gpc^{-3}}$ at $m_{\rm PBH} \approx 2.5 M_\odot$~\cite{Abbott:2020uma} and of $\tau \approx 50 {\rm yr^{-1} Gpc^{-3}} $ at masses between $ 10 M_\odot$ and $50 M_\odot$~\cite{LIGOScientific:2018jsj}. The boost in primordial BH formation at the time of the QCD transition will induce a peak at the solar mass scale in any primordial BH model with an extended mass function~\cite{Byrnes:2018clq,Carr:2019kxo}.  If one normalises the merging rates at the peak with the LIGO/Virgo rates in the solar mass range, then one obtains an upper bound on the rate distribution
    \be  \label{eq:ratescatpure2}
    \frac{d\tau}{d(\ln m_1) d(\ln m_2)} \approx 4 \times 10^3 \times \rho(m_1) \rho(m_2) \frac{(m_1 + m_2)^{10/7}}{(m_1 m_2)^{5/7}} \rm{yr^{-1}Gpc^{-3}} \,,
    \ee
    while being agnostic about the total primordial BH abundance, $f_{\rm PBH}$. Then, like for primordial binaries, one can obtain an upper limit on the merging rate for equal-mass sub-solar binaries in halos. Assuming $\rho(m_{\rm PBH}) \approx 0.01$, as suggested by microlensing surveys, one gets
    \be
    \tau \approx 1.2 \, \rm{yr^{-1}Gpc^{-3}} \,,
    \ee
    independently of the primordial BH mass. This is of the same order than the rates obtained with the theoretical prescriptions leading to Eq.~\eqref{eq:capturerate}, for a monochromatic distribution. One thus expects that Eq.~\eqref{eq:ratescatpure2} is a good approximation for a broad variety of primordial BH scenarios, for both sharp and wide mass distributions. One then obtains the required experimental strain sensitivity to detect one event per year,
    \be
    h_{\rm max}  \approx 2.8 \times 10^{-23}  \left(\frac{m_{\rm PBH}}{M_\odot}\right) \approx 6.1 \times 10^{-20} \left( \frac{\rm Hz}{f} \right) \,.
    \ee
     This GW signal is therefore typically lower than for primordial BH binaries formed in the early Universe.  However, it is still debated which of the binary formation channel is dominant, especially if $f_{\rm PBH} \gtrsim 0.1$, i.e. if primordial BHs explain a significant or even the totality of the dark matter in the Universe.
\end{enumerate}
While in the literature it is often assumed for simplicity that primordial BHs have a monochromatic mass function, this is not expected in a realistic scenario. Even in the limiting case of BH formation due to a sharp peak in the primordial power spectrum, these primordial BHs would have a relatively broad distribution, due to effects related to the critical collapse~\cite{Musco:2008hv,Musco:2012au}.  For this reason and to be generic, we have also estimated in the above paragraph the rate distribution for the two possible binary formation mechanisms, without specifying the primordial BH mass function.

\subsubsection{Exotic compact objects }
\label{sec:ECOs}
Beyond the very well-known astrophysical compact objects, namely BHs and neutron stars, there are several candidates for stable (or long-lived) exotic compact objects that are composed of beyond the Standard Model particles~\cite{Giudice_2016}. For instance, they can be composed of beyond the Standard Model fermions, such as the gravitino in supergravity theories, giving rise to gravitino stars~\cite{Narain_2006}. Exotic compact objects can also be composed of bosons, such as moduli in string compactifications and supersymmetric theories~\cite{Krippendorf_2018}. Depending on the mechanism that makes the compact object stable (or long-lived), scalar field exotic compact objects have specific names such as Q-balls, boson stars, oscillatons, oscillons. There are also more exotic possibilities, such as gravastars~\cite{mazur2001gravitational}. Exotic compact objects can form binaries and emit GWs in the same way as BH and neutron star binaries do. During the early inspiral phase, the frequency of the emitted GWs is twice the orbital frequency. At the ISCO, the frequency for a binary system of two exotic compact objects with mass $M$ and radius $R$ is given by~\cite{Giudice_2016}
\begin{equation}
f_{\rm ISCO} = \frac{1}{6 \sqrt{3} \pi} \frac{C^{3/2}}{G M} \simeq C^{3/2} \left(\frac{6 \times 10^{-3} \, M_\odot}{M}\right) \, 10^6 \, \text{Hz} \,,
\label{eq:fISCO-ECO}
\end{equation}
where $C = G M/R$ is the compactness of the exotic compact object. This expression is only slightly modified for a boson star binary with two different values of the masses. Note that for a BH the radius is given by the Schwarzschild radius $R_{\text{S}} = 2 G M$, therefore $C = 1/2$ is the maximum attainable value for the compactness.

The GW strain for a boson star binary formed by equal mass objects $M$ can be estimated as done in Sec.~\ref{sec:PBHmergers}
\begin{equation}
h_0 \simeq 1.72 \times 10^{-20}\, C \, \left(\frac{M}{M_\odot}\right) \, \left(\frac{\text{Mpc}}{D}\right) \,,
\label{eq:h0-ECO}
\end{equation}
where $D$ is the distance between the source and the observer. The exact waveform produced by the merger of two exotic compact objects is in general different from that of BHs and neutron stars and depends on its microphysics details~\cite{Giudice_2016, Palenzuela_2017}.\footnote{See however~\cite{Helfer:2018vtq} for more details on the initial conditions.} Hence, the detection of GWs from an exotic compact object merger would give further valuable information about beyond the Standard Model physics.

\subsubsection{Black hole superradiance}
\label{sec:Superradiance}

This section focuses on GW emission from clouds of axions or axion-like particles created by the gravitational superradiance of BHs~\cite{Ternov:1978gq,Zouros:1979iw,Arvanitaki:2009fg,Arvanitaki:2010sy,arvanitaki:2016gw,aggarwal2020searching,Detweiler:1980uk,Yoshino:2013ofa,Arvanitaki:2014wva,Brito:2014wla,Brito:2015oca}.
Superradiance is an enhanced radiation process that is associated with bosonic fields around rotating objects with dissipation. The event horizon of a spinning BH is one such example that provides conditions particularly suitable for superradiance~\cite{Arvanitaki:2014wva}.
When the axion Compton wavelength, determined by the axion mass $m_a$, is about the size of the BH,
\begin{equation}
\label{eq:maMBH}
m_a \sim \left( \frac{ M_\odot}{M_{\rm BH} } \right) \, 10^{-10}\, \text{eV} \,,
\end{equation}
the axions can accumulate outside the BH event horizon and outside the ergosphere efficiently. 
The BH forms a gravitationally bound `atom' with the axions, with different atomic `levels' occupied by exponentially large numbers of axions.

The primary candidate for GWs at high-frequencies is the axion annihilation  process (${\bf a}  +  {\bf a} \rightarrow h$), with frequencies of around 100 kHz for $M_{\rm BH} \gtrsim M_\odot$.The GW frequency emitted by this process is twice the Compton frequency of the axion, i.e.
\begin{equation}
\label{eq:fsprannihilation}
f = 2 \left( \frac{m_a}{10^{-9} \, \text{eV}} \right) \, 10^{6} \, \text{Hz} \,.
\end{equation}
 The expected GW strain for this process is roughly \cite{arvanitaki:2016gw} 
\begin{equation}
    h_0 \sim 10^{-19}\left(\frac{\alpha}{l}\right)\epsilon \left(\frac{10 \, \mathrm{kpc}}{D}\right)\left(\frac{M_\mathrm{BH}}{2 M_\odot}\right) \,,
\label{eq:strain_axion_annihilation}
\end{equation}
where $\alpha = G M_\mathrm{BH} \, m_a$, $l$ is the orbital angular momentum number of the axions that decay, $D$ is the distance from the observer and $\epsilon < 10^{-3}$ denotes the fraction the BH mass accumulated in the axion cloud. {The superradiance condition constrains $\alpha/l < 0.5$~\cite{Arvanitaki:2010sy}.} See Refs. \cite{Brito:2014wla,aggarwal2020searching} for more recent calculations of GW strain from BH superradiance leading to axion annihilations.

If lighter spinning BHs exist (see Sec.~\ref{sec:PBHmergers}), then this process can produce GWs of even higher frequencies.  Note that this GW signal is predicted to be monochromatic and coherent~\cite{Arvanitaki:2014wva}, rendering it quite distinct from any other astrophysical or cosmological sources discussed here.

For heavier BHs and lighter axions, this process can produce GWs in the LIGO/VIRGO band. Additionally, these bosons created in BH superradiance may also undergo transition from one level to another, emitting a graviton in the process. Both these processes would  produce GWs of lower frequencies and could be detectable in LIGO-VIRGO and LISA, e.g. Refs.~\cite{Brito+2017,Zhu:2020tht,Tsukada:2018mbp}.

Finally, recently it has been postulated that axions might also decay into gravitons (${\bf a}\to h \, h$)~\cite{Sun:2020gem}. In such a process, the GW frequency would be half of the axion Compton frequency, i.e.
\begin{equation}
\label{eq:fsprdecay}
f = \frac{1}{2}\left( \frac{m_a}{10^{-9} \, \text{eV}} \right) \, 10^{6} \, \text{Hz} \,.
\end{equation}
The corresponding strain of the coherent signal has been calculated in Ref.~\cite{Sun:2020gem} to be
\begin{align}\label{GWhg1}
  h_0  \sim 10^{-24} \left(\frac{1 \, \text{MHz}}{f}\right)  \left(\frac{\epsilon \, M_\text{BH}}{ 10^{-7} M_{\odot}} \right)^{1/2} \left(\frac{10 \, \text{kpc}}{D}\right) \,,
     \end{align}
where $\epsilon < 10^{-3}$ denotes the fraction the BH mass accumulated in the axion cloud.

\subsection{Early Universe \label{sec:earlyU} }

We now turn to cosmological sources emitting GWs at cosmological distances, i.e.\ in the early Universe. For a summary of these sources see Fig.~\ref{fig:hc_stochastic_summary} and Tab.~\ref{tab:summary-stochastic} in App.~\ref{sec:SummaryTable}. In this case, the source is associated to an event in our cosmological history, triggered e.g.\ by the decreasing temperature $T$ of the thermal bath, and typically occurs everywhere in the Universe at (approximately) the same time. This results in a stochastic background of GWs which is a superposition of GWs with different wave vectors. The total energy density of a GW background $\rho_{\rm GW} \equiv \int d\log k \left(d\rho_{\rm GW} / d\log k\right)$, with characteristic wavelengths well inside the horizon, decays as relativistic degrees of freedom with the expansion of the Universe, i.e.~as $\rho_{\rm GW}\propto a^{-4}$. This implies that a GW background acts as an additional radiation field contributing to the background expansion rate of the Universe. Observables that can probe the background evolution of the Universe at some particular moment of its history, can therefore be used to constrain $\rho_{\rm GW}$ at such moments. In particular, two events in cosmic history yield a precise measurement of the expansion rate of the Universe: BBN and photon decoupling of the CMB. An upper bound on the total energy density of a GW background present at the time of BBN and CMB decoupling can be therefore derived from the constraint on the amount of radiation tolerated at those cosmic epochs, when the Universe had a temperature of $T_{\rm BBN} \sim 0.1$ MeV and $T_{\rm CMB} \sim 0.3$ eV, respectively.

A constraint on the presence of `extra' radiation is usually expressed in terms of an effective number of neutrinos species $N_{\rm eff}$ after electron-positron annihilation. After electron-positron annihilation, the total number of Standard Model relativistic degrees of freedom was $g_*(T < T_{e^+e^-}) = 2 + \frac{7}{4}\,N_{\rm eff} \left(\frac{4}{11}\right)^{4/3}$, with $N_{\rm eff}=3.046$. As the radiation energy density for thermal degrees of freedom in the Universe is given by $\rho_{\rm rad} = \frac{\pi^2}{30}g_{*}(T)T^4$, an extra amount of radiation can be parametrized by $\Delta N_{\rm eff}$ extra neutrino species\footnote{This is independent of whether the extra radiation is in a thermal state or not, as this is only a parametrization of the total energy density of the extra component, independently of its spectrum.}, as $\Delta \rho_\mathrm{rad} =  \frac{\pi^2}{30}\, \frac{7}{4}\, \left(\frac{4}{11}\right)^{4/3} \Delta N_{\rm eff} \, T^4$. An upper bound on the extra radiation can thus be seen as an upper bound on $\Delta N_{\rm eff}$. Since the energy density in GW must satisfy $\rho_{\rm GW}(T) \leq \Delta \rho_\mathrm{rad}(T)$, we obtain $\left(\frac{\rho_\mathrm{GW}}{\rho_{\gamma}}\right)_{T = \mathrm{MeV}} \leq \frac{7}{8}\, \left(\frac{4}{11}\right)^{4/3} \Delta N_{\rm eff}$, with $\rho_\gamma$ denoting the energy density in photons. Writing the fraction of GW energy density today\footnote{{We write the current value of the Hubble parameter as $H_0 = h_H \times 100 \, \text{km} \, \text{sec}^{-1} \, \text{Mpc}^{-1}$. Early Universe and late time observations report slightly different values for $h_H$, see~\cite{Bernal:2016gxb} for a discussion. For all our purposes, we will assume $h_H = 0.7$ when needed.}} as $\left(\frac{\rho_{\rm GW} \, h_H^2}{\rho_c}\right)_0 = \Omega_{\text{rad},0} \, h_H^2 \, \left(\frac{g_S(T_0)}{g_S(T)}\right)^{4/3}\frac{\rho_{\rm GW}(T)}{\rho_\gamma(T)}$, we obtain a constraint on the redshifted GW energy density today, in terms of the number of extra neutrino species~\cite{Caprini:2018mtu}
\begin{eqnarray}
\label{eq:ConsRhoBBN}
\left(\frac{\rho_{\rm GW} \, h_H^2}{\rho_c}\right)_0 \leq  \Omega_{\text{rad},0} \, h_H^2 \times \frac{7}{8}\, \left(\frac{4}{11}\right)^{4/3} \Delta N_{\rm eff}= 5.6 \times 10^{-6} \,
\Delta N_{\rm eff}\,,
\end{eqnarray}
where we have inserted $\Omega_{\text{rad},0} \, h_H^2 = \left(\rho_{\gamma}/\rho_c\right)_0 \, h_H^2 = 2.47 \times 10^{-5}$.
We recall that the above bound applies only to the total GW energy density, integrated over wavelengths way inside the Hubble radius (for super-horizon wavelengths, tensor modes do not propagate as a wave, and hence they do not affect the expansion rate of the Universe). Except for GW spectra with a very narrow peak of width $\Delta f \ll f$, the bound can be interpreted as a bound on the amplitude of a GW spectrum, $\Omega_{\text{GW},0}(f) \, h_H^2 \lesssim 5.6 \times 10^{-6} \Delta N_{\rm eff}$, over a wide frequency range.  The bound obviously applies only to GW backgrounds that are present before the physical mechanism (BBN or CMB decoupling) considered to infer the constraint on $N_{\rm eff}$ takes place.

Constraints on $N_{\rm eff}$ can be placed by BBN alone, and/or in combinations with CMB data. In particular, Ref.~\cite{Cyburt:2015mya} finds $\Delta N_{\rm eff} < 0.2$ at 95\% confidence level. Eq.~\eqref{eq:ConsRhoBBN} then gives
\begin{eqnarray}\label{eq:BBNbound}
\left(\frac{\rho_{\rm GW} \, h_H^2}{\rho_c}\right)_0 < 1.12 \times 10^{-6}\,,
\label{eq:BBN}
\end{eqnarray}
for a stochastic GW background produced before BBN, with wavelengths inside the Hubble radius at the onset of BBN, corresponding to present-day frequencies $f \geq 1.5 \times 10^{-12}$ Hz.
A similar bound can also be obtained from constraints on the Hubble rate at CMB decoupling~\cite{Smith:2006nka,Sendra:2012wh,Pagano:2015hma, Clarke:2020bil}. This translates into an upper bound on the amount of GWs, which extends to a greater frequency range than the BBN bound, down to $f \gtrsim 10^{-15}$ Hz.

Since high-frequency GWs carry a lot of energy, $\Omega_\text{GW} \propto f^3 \, S_h$, these bounds pose severe constraints on possible cosmological sources of high-frequency GWs.

\subsubsection{Inflation}
\label{sec:Inflation}
Under the standard assumption of scale invariance, the amplitude of the GWs produced during inflation is too small ($\Omega_{\rm GW, 0} \lesssim 10^{-16}$) to be observable with current technology.\footnote{However, note that the proposed space-borne detectors Big Bang Observer (BBO)~\cite{BBO} and DECI hertz Interferometer Gravitational wave Observatory (DECIGO)~\cite{Seto:2001qf} may reach the necessary sensitivity, assuming that astrophysical GW foregrounds will be able to be substracted to this accuracy}.

Various inflationary mechanisms have been studied in the literature that can produce a significantly blue-tilted GW signal (i.e.\ increasing towards higher frequency), or a localized bump at some given (momentum) scale, with a potentially visible amplitude. A number of these mechanisms have been explored in~\cite{Bartolo:2016ami} with a focus on the LISA experiment, and therefore on a GW signal in the mHz range. However, these mechanisms can be easily extended to higher frequencies. Assuming an approximately constant Hubble $H$ parameter during inflation, a GW generated $N$ Hubble times (e-folds) before the end of inflation with frequency $H$ is redshifted to frequency $f$ today, with
\begin{equation}
\ln \left[\frac{f}{10^{-18} {\rm Hz}}\right] \simeq N_{\rm CMB}-N \,,
\label{eqfN}
\end{equation}
where $ N_{\rm CMB}$ is the number of e-folds at which the CMB modes exited the horizon. The numerical value of $N_{\rm CMB}$ depends logarithmically on the energy scale of inflation, which is bounded from above by the upper bound on the tensor-to-scalar ratio~\cite{Akrami:2018odb}, $H \lesssim 6 \times 10^{13}$~GeV. Saturating this bounds implies $N_{\rm CMB} \simeq 60$, and a peak at $f = 1 \; {\rm MHz}$ then corresponds to the $N = 4.7$, while the LIGO frequency $f_{\rm LIGO} = {\mathcal{O}} \left( 10^2 \, {\rm Hz} \right)$ corresponds to about $N = 14$. Such late stages of inflation are not accessible by electromagnetic probes.

Ref.~\cite{Bartolo:2016ami} discusses three broad categories: the presence of extra fields that are amplified in the later stages of inflation (so to affect only scales much smaller than the CMB ones); GW production in the effective field theory framework of broken spatial reparametrizations and GWs sourced by (large) scalar perturbations.
In the following we will briefly summarize these three cases.

\paragraph{Extra-species}
Several mechanisms of particle production during inflation, with a consequent GW amplification, have been considered in the recent literature. Here, for definiteness, we discuss a specific mechanism in which a pseudo-scalar inflation $\phi$ produces gauge fields via an axionic coupling $\frac{\phi }{4 f_a} F {\tilde F}$,  where $F_{\mu \nu}$ is the gauge field strength,  ${\tilde F}_{\mu \nu}$ its dual, and $f_a$ is the axion decay constant. The motion of the inflaton results in a large amplification of one of the two gauge field helicities. The produced gauge quanta in turn generate inflaton perturbations and GW via $2 \to 1$ processes~\cite{Barnaby:2010vf,Sorbo:2011rz}. In particular, the spectrum of the sourced GWs is~\cite{Barnaby:2010vf}
\begin{equation}
\Omega_{\rm GW, 0} \simeq 3.6 \cdot 10^{-9} \, \Omega_{\text{rad},0} \frac{H^4}{M_p^4} \, \frac{{\rm e}^{4 \pi \xi}}{\xi^6}\,, \qquad
\xi \equiv \frac{\dot{\phi}}{2 f_a H} \,,
\label{PGW-frequency}
\end{equation}
where $H$ is the Hubble rate. In this relation, $H$ and $\dot{\phi}$ are evaluated when a given mode exits the horizon, and therefore the spectrum in Eq.~\eqref{PGW-frequency} is in general scale-dependent. In particular, in the $\xi \gg 1$ regime, the GW amplitude grows exponentially with  the speed of the inflaton, which in turn typically increases over the course of inflation in single-field inflation models. As a consequence, the spectrum in Eq.~\eqref{PGW-frequency} is naturally blue. The growth of $\xi$ is limited by the backreaction of the gauge fields on the inflaton. {Within the limits of a perturbative description, $\xi \lesssim 4.7$~\cite{Peloso:2016gqs}, GW amplitudes of $\Omega_{\text{GW},0} \simeq 10^{-10}$ can be obtained.} Refs.~\cite{Domcke:2016bkh,Garcia-Bellido:2016dkw} explored the resulting spectrum for several inflaton potentials. In particular hill-top potentials are characterized by a very small speed close to the top (that is mapped to the early stages of observable inflation), and by a sudden increase of the speed at the very end of inflation. Interestingly, hill-top type potentials are naturally present~\cite{Peloso:2015dsa} in models of multiple axions such as aligned axion inflation~\cite{Kim:2004rp}.

\paragraph{Effective field theory spatial reparametrizations}
The modification of  the theory of gravity which underlines the inflationary physics can give rise to an extra production of GWs with a large amplitude (and blue tilt) rendering it accessible to high-frequency GW experiments. From the theoretical point of view, the effective field theory approach~\cite{Cheung:2007st} represents a powerful tool to provide a clear description of the relevant degrees of freedom at the energy scale of interest exploiting the power of symmetries and gives an accurate prediction of observational quantities.
In the standard single-field effective field theory of inflation ~\cite{Cheung:2007st} only time-translation symmetry ($t\rightarrow t+\xi_{0}$) is broken according to the cosmological background expansion during inflation. However when space-reparameterization symmetry  ($x_{i}\rightarrow x_{i}+\xi_{i}$) is also broken~\cite{Bartolo:2015qvr,Graef:2015ova}, scalar and tensors (GWs) acquire interesting features. In particular tensors can acquire a non-trivial mass $m_h$ and sound speed $c_T$, making them potential targets for high-frequency detectors since in this case the spectrum gets enhanced on small scales. At the quadratic level in perturbations, in an effective field theory approach, the action for graviton fluctuations $h_{ij}$ around a conformally flat
Friedmann-Lema$\hat{\text{i}}$tre-Robertson-Walker background can be expressed
as in~\cite{Cannone:2014uqa, Bartolo:2015qvr,Ricciardone:2016lym}:
\begin{equation} \label{sol-qac}
\mathcal{L}_{h}=\frac{M_p^2}{8}\,\left[\dot h_{ij}^2 -\frac{c_T^2(t)}{a^2} \,\left( \partial_l h_{ij}\right)^2
-m_h^2(t)\,h_{ij}^2\right]\,.
\end{equation}
The corresponding tensor power spectrum and its related spectral tilt are:
\begin{equation}
{\cal P}_T\,=\,\frac{2\,H^2}{\pi^2\,M_p^2\,c_T^3}\,\left( \frac{k}{k_*}\right)^{n_T}\,,\quad \quad\quad \quad n_T\,=\,-2 \epsilon+\frac{2}{3}
\frac{m_{h}^2}{H^2}\,.
\label{nteft}
\end{equation}
Hence, if the quantity $m_{h}/H$ is sufficiently large,  we can get a blue tensor spectrum with no need to violate the null energy condition in the early Universe. Consequently $\Omega_{\text{GW},0} \sim \Omega_{\text{rad},0} {\cal P}_T$ is enhanced at high-frequencies, making it a potential target for high-frequency GW detectors. The upper bound at on the spectrum at high-frequencies is set by the observational BBN and CMB bounds, see Eq.~\eqref{eq:BBN}. This scenario shows how GW detectors at high-frequency might be useful to test the modification of gravity at very high-energy scales.

\paragraph{Second-order GW production from primordial scalar fluctuations}
In homogeneous and isotropic backgrounds, scalar, vector and tensor fluctuation modes decouple from each other at first order in perturbation theory. These modes can however source each other non-linearly, starting from second order. In particular, density perturbations can produce  `induced' (or `secondary') GWs through a $\zeta+\zeta \to h$ process \cite{Mollerach:2003nq,Ananda:2006af,Baumann:2007zm} (see also \cite{Kohri:2018awv,Espinosa:2018eve, Braglia:2020eai}). This production, which simply involves only gravity, is mostly effective when the modes re-enter the horizon after inflation. (Second order GWs are also produced in an early matter era, \cite{Inomata:2019ivs,Inomata:2019zqy})
The amplitude of this signal is quadratic in the scalar perturbations, and scale-invariant ${\mathcal{O}} \left( 10^{-5} \right)$  perturbations, as measured on large scales by the CMB, result in unobservable GWs due to too small amplitude. On the other hand, if the spectrum of scalar perturbations produced during inflation has a localized bump at some given scale (significantly smaller than the scales of CMB and the large scale structure), as required e.g.\ to obtain a sizable primordial BH abundance of some specific given mass, the height of the bump could be sufficiently high to produce a noticeable amount of GWs~\cite{Inomata:2016rbd,Garcia-Bellido:2017aan,Bartolo:2018rku}. The non-detection of the stochastic GW background can also be used to constrain fluctuations \cite{Byrnes:2018txb,Inomata:2018epa}. The induced GWs have a frequency $f_*$ parametrically equal to the momentum $k_*$ and can hence be related to the e-fold $N$ of horizon exit of the scalar perturbation through Eq.~\eqref{eqfN}.

The precise amount of produced GWs depends on the statistics of the scalar perturbations~\cite{Nakama:2016gzw,Garcia-Bellido:2017aan,Cai:2018dig,Unal:2018yaa}. A reasonable estimate is however obtained by simply looking at the scalar two-point function,
\begin{equation}
P_{h}^{\rm ind}  \propto \langle h^2 \rangle  \propto \langle \zeta^4 \rangle \propto P_\zeta^2 \,,
\end{equation}
where $P_\zeta$ is the power spectrum (two-point function) of the gauge invariant scalar density fluctuations such that $\langle \zeta_{\bf k} \, \zeta_{\bf k'} \rangle \propto  \frac{\delta ({\bf k+k'})}{k^3} \, P_\zeta (k)$. From this relation, the present value of the induced stochastic GW background is given by
\begin{equation}
\Omega_{\text{GW},0} \sim  \Omega_{\text{rad},0} \,  P_\zeta^2 \,.
\end{equation}

At the largest scales of our observable Universe, the density fluctuations are measured as $P_\zeta \simeq 2 \cdot 10^{-9}$, resulting in $\Omega_{\text{GW},0} \sim  {\cal O}(10^{-22})$. Primordial BH limits are compatible with $P_\zeta$ as large as $\lesssim 10^{-2.5}$ at some (momentum) scale $k_*$, in which case $\Omega_{\text{GW},0} \sim  {\cal O} (10^{-9})$.

\subsubsection{(P)reheating}
\label{sec:Preheating}

Preheating is an out-of-equilibrium production of particles due to non-perturbative effects \cite{Traschen:1990sw,Kofman:1994rk,Shtanov:1994ce,Kaiser:1995fb,Khlebnikov:1996mc,Prokopec:1996rr,Kaiser:1997mp,Kofman:1997yn,Greene:1997fu,Kaiser:1997hg}, which takes place after inflation in many models of particle physics (see \cite{Allahverdi:2010xz,Amin:2014eta,Lozanov:2019jxc} for reviews).  After inflation, the interactions between the different fields may generate non-adiabatic time-dependent terms in the field equations of motion, which can give rise to an exponential growth of the field modes within certain bands of momenta. The field gradients generated during this stage can be an important source of primordial GWs, with the specific features of the GW spectra depending strongly on the considered scenario, see e.g.~\cite{Khlebnikov:1997di,GarciaBellido:1998gm,Easther:2006gt,Easther:2006vd,GarciaBellido:2007dg,GarciaBellido:2007af,Dufaux:2007pt,Dufaux:2008dn,Figueroa:2011ye}. If instabilities are caused by the field's own self-interactions, we refer to it as \textit{self-resonance}, a scenario which will be discussed in more detail below. Here we consider instead a multi-field preheating scenario, in which a significant fraction of energy is successfully transferred from the inflationary sector to other fields.

For illustrative process, let us consider a two-field scenario, in which the post-inflationary oscillations of the inflaton excite a secondary massless species. More specifically, let us consider an inflaton with power-law potential $V(\phi) = \frac{1}{p} \lambda \mu^{4-p} |\phi|^p$, where $\lambda$ is a dimensionless coefficient, $\mu$ is a mass scale, and $p\geq 2$. Let us also define $t_{\star}$ as the time when inflation ends. For $t \gtrsim t_{\star}$, the inflaton oscillates with time-dependent frequency $\Omega_{\rm osc} \equiv \omega_{\star} (t /t_{\star})^{1- 2/p}$, where $\omega_{\star} \equiv \sqrt{\lambda} \mu^{(2 - p/2)} \phi_{\star}^{(p/2 - 1)}$ and $\phi_{\star} \equiv \phi (t_{\star})$ \cite{Turner:1983he}. Let us now include a quadratic interaction term $g^2 \phi^2 \chi^2$ between the inflaton and a secondary massless scalar field $\chi$, where $g$ is a dimensionless coupling constant. In this case, the driving post-inflationary particle production mechanism is parametric resonance \cite{Kofman:1994rk,Kofman:1997yn,Greene:1997fu}. In particular, if the so-called \textit{resonance parameter} $q_{\star} \equiv g^2 \phi_{\star}^2 /\omega_{\star}^2$ obeys $q_{\star} \gtrsim 1$, the secondary field gets excited through a process of broad resonance, and the amplitude of the field modes grows exponentially inside a Bose-sphere of radius $k \lesssim k_{\star} \sim q_*^{1/4} \omega_{\star}$. The GW spectrum produced during this process has a peak at approximately the frequency and amplitude \cite{Figueroa:2017vfa},
\begin{subequations}
    \begin{gather}
        f \simeq  8 \cdot 10^{9} \left( \frac{\omega_{\star}}{{\rho}_{\star}^{1/4}}\right)\, \epsilon_{\star}^{\frac{1}{4}} q_{\star}^{\frac{1}{4} + \eta}~{\rm Hz} \label{eq:preheating-1} \,,\\
        \Omega_{\text{GW},0}(f) \simeq \mathcal{O}(10^{-9}) \times \epsilon_{\star} \, \mathcal{C} \frac{\omega_{\star}^{6}}{\rho_{\star} M_p^{2}}\,q_{\star}^{-\frac{1}{2}+\delta} \label{eq:preheating-2} \,,
    \end{gather}
\end{subequations}
where $\rho_{\star}$ is the energy density at time $t=t_{\star}$, $\eta$ and $\delta$ are two parameters that account for non-linear effects, and $\mathcal{C}$ is a constant that characterizes the strength of the resonance. The factor $\epsilon_{\star} \equiv (a_{\star} /a_{\rm RD})^{1 - 3 w}$ parametrizes the period between the end of inflation and the onset of the radiation dominated stage with a transitory effective equation of state $w$. If non-linear effects are ignored, the frequency and amplitude scale as $f \sim q_{\star}^{1/4}$ and $\Omega_{\text{GW},0} \sim q_{\star}^{-1/2}$ respectively.

The values for $\mathcal{C}$, $\eta$, and $\delta$, can be determined for specific preheating models with classical lattice simulations. For chaotic inflation with quadratic potential $V(\phi) \propto \phi^2$, one finds a frequency in the range $f \simeq (10^8 - 10^9) \, {\rm Hz}$ and $\Omega_{\text{GW},0} \simeq (10^{-12} - 10^{-11})$ for resonance parameters $q_{\star} \in (10^4,10^6)$ (assuming $\epsilon_{\star} = 1$). On the other hand, for the quartic potential $V(\phi) \propto \phi^4$, one gets $f \simeq (10^7 - 10^8)  \,  {\rm Hz}$  and $\Omega_{\text{GW}, 0} \simeq (10^{-13} - 10^{-11}) $ in the range $q_{\star} \in (1,10^4)$. The GW spectrum in the quartic case also features additional peaks, see~\cite{Figueroa:2017vfa} for more details.

GWs can also be strongly produced if the species of the fields involved is different, or when the resonant phenomena driving preheating is different than parametric resonance. For example, GWs can be produced during the out-of-equilibrium excitation of fermions after inflation, for both spin-1/2 \cite{Enqvist:2012im,Figueroa:2013vif,Figueroa:2014aya} and spin-3/2 \cite{Benakli:2018xdi} fields. Similarly, GWs can also be generated when the decay products are (Abelian and non-Abelian) gauge fields. For example, the gauge fields can be coupled to a complex scalar field via a covariant derivative like in \cite{Dufaux:2010cf,Figueroa:2016ojl,Tranberg:2017lrx}, or to a pseudo-scalar field via an axial coupling as in~\cite{Adshead:2018doq,Adshead:2019igv,Adshead:2019lbr}. Preheating can be remarkably efficient in the second case, and the GW amplitude can scale up to $\Omega_{\rm GW} \sim \mathcal{O} \left(10^{-6} - 10^{-7}\right)$ for certain coupling strengths, see \cite{Adshead:2019igv,Adshead:2019lbr} for more details. Production of GWs during preheating with non-minimal couplings to the scalar curvature has also been explored in \cite{Fu:2017ero}. Finally, the stochastic background of GWs from preheating may develop anisotropies if the inflaton is coupled to a secondary light scalar field, see \cite{Bethke:2013aba, Bethke:2013vca}.

\paragraph{Oscillon Production.}

Oscillons are long-lived compact objects~\cite{Gleiser:1993pt} that can be formed in the early Universe in a variety of post-inflationary scenarios which involve a preheating-like phase~\cite{Amin:2010jq, Amin:2010dc, Amin:2011hj, Zhou:2013tsa, Amin:2013ika, Lozanov:2014zfa, Antusch:2015vna, Antusch:2015ziz, Antusch:2016con, Antusch:2017flz, Antusch:2017vga, Lozanov:2017hjm, Amin:2018xfe, Antusch:2019qrr, Sang:2019ndv, Lozanov:2019ylm, Fodor:2019ftc, Hiramatsu:2020obh}. Their dynamics is a possible source of GW production. Oscillons are pseudo-solitonic solutions of real scalar field theories: their existence is due to attractive self-interactions of the scalar field that balance the outward pressure.\footnote{If the scalar field is complex and the potential features a global $U(1)$ symmetry, non-topological solitons like Q-balls~\cite{Coleman:1985ki} can be formed during the post-inflationary stage, giving rise to  similar GW signatures~\cite{Chiba:2009zu}.} The real scalar field self-interactions are attractive if the scalar potential is shallower than quadratic at least on one side with respect to the minimum. Oscillons can be thought off as bubbles in which the scalar field is undergoing large oscillations that probe the non-linear part of the potential, while outside the scalar field is oscillating with a very small amplitude around the minimum of the potential.

As discussed in the previous section, during preheating the quantum fluctuations of the scalar field are amplified due to a resonance process. The Universe ends up in a very inhomogeneous phase in which the inflaton (or any other scalar field that produces preheating) is fragmented and there are large fluctuations in the energy density. At this point, if the field is subject to attractive self-interactions, the inhomogeneities can clump and form oscillons. While clumping oscillons deviate significantly from being spherically symmetric, therefore their dynamics produce GWs. After many oscillations of the scalar field they tend to become spherically symmetric and GW production stops. However, during their entire lifetime oscillons can produce GWs also due to the interactions and collisions among each other~\cite{Helfer:2018vtq}. Oscillons are very long-lived: their lifetime is model-dependent but typically $\gtrsim 10^4 /m$~\cite{Gleiser:2008ty, Amin:2010jq, Amin:2010dc, Amin:2011hj, Salmi:2012ta, Saffin:2014yka, Antusch:2019qrr, Gleiser:2019rvw, Zhang:2020bec}, where $m$ is the mass of the scalar field. Oscillons eventually decay through classical~\cite{Segur:1987mg} or quantum radiation~\cite{Hertzberg:2010yz}.

The peak of the GW spectrum at production is centered slightly below the value of the mass of the field, that typically correspond to a frequency today well above the LIGO range\footnote{See however~\cite{Antusch:2016con, Liu:2017hua, Kitajima:2018zco} for models that lead to a GW peak at lower frequencies.}~\cite{Zhou:2013tsa, Antusch:2017flz, Lozanov:2019ylm}. In a typical situation, an oscillating massive scalar field forming oscillons quickly comes to dominate the energy density of the Universe until the perturbative decay of the field itself. For the simplest case of a gravitationally coupled massive field that starts oscillating at $H \simeq m$ and decays at $H \sim m^3/M_p^2$) the frequency today can be estimated as
\begin{equation}
\label{eq:OscillonFrequency}
f \simeq X \left(\frac{m}{10^{12} \, \text{GeV}}\right)^{5/6} \, 10^6 \, \text{Hz} \,,
\end{equation}
where the factor $X$ which is typically in the range $X \simeq (10-10^3)$ is due to the unknown precise time of GW production and can be obtained in concrete models through lattice simulations: the equality would hold if GWs were produced immediately when the scalar field starts oscillating.\footnote{This rough estimate assumes that the field starts oscillating when $H \simeq m$. Since the potential contains self-interactions, assuming that the field starts at rest, the actual requirement for the start of the oscillations is $V''(\phi_{\text{in}}) \sim H$, where $\phi_{\rm in}$ is the initial value of the field. Please note that if the field is the inflaton, the initial conditions are different from those assumed in Eq.~\eqref{eq:OscillonFrequency} and therefore this estimate does not necessarily hold, see e.g.~\cite{Antusch:2016con}.} On the other hand, the later GWs are produced, the less the frequency is red-shifted and the larger is $X$. The maximum value of today's amplitude for these processes, inferred from numerical simulations, is in the range $\Omega_{\text{GW},0} \simeq (10^{-13} - 10^{-10})$~\cite{Antusch:2016con, Antusch:2017flz, Amin:2018xfe}, see~\cite{Dufaux:2007pt} for a discussion on how to compute the GW amplitude.

Depending on the model, gravitational effects can become important and play a crucial role for the existence/stability of the solution~\cite{Seidel:1991zh}. In particular the requirement that the potential must be shallower than quadratic is no longer necessary, as the attractive force is provided by gravity~\cite{UrenaLopez:2002gx}. In this case oscillons are equivalent to oscillatons, see Sec.~\ref{sec:ECOs}, and can give rise to interesting additional effects, such as the collapse to BHs~\cite{Muia:2019coe, Giblin:2019nuv, Kou:2019bbc, Nazari:2020fmk}.


\subsubsection{Cosmic gravitational microwave background}
\label{sec:CGMB}

The hot thermal plasma of the early Universe acts as a source of GWs, which, similarly to the relic photons of the CMB, peak in the $\sim 100$~GHz range today. The spectrum of this signal is determined by the particle content and the maximum temperature $T_{\rm max}$ reached by the thermal plasma in the Universe history~\cite{Ghiglieri:2015nfa, Ghiglieri:2020mhm, Ringwald:2020ist}. Ignoring the dependence on the number of relativistic degrees of freedom, the energy density in GWs per logarithmic frequency interval can then be written as
 \begin{align}
\Omega_{\rm GW,0}(f) \simeq \frac{1440 	\sqrt{10}}{2 \pi^2} \Omega_{{\rm rad}, 0} \frac{f^3}{T_0^3} \frac{T_{\rm max}}{M_p} \hat{\eta} \left(T_{\rm max}, 2 \pi f/T_0\right)\,,
  \label{eq:CGMB}
 \end{align}
where $T_0$ is the temperature of the CMB today, while $\hat{\eta}(T_{\rm max}, 2 \pi f/T_0)$ encodes the sources of GW production in the thermal plasma: it is dominated by long range hydrodynamic fluctuations at $2 \pi f<T_0$ and by quasi-particle excitations in the plasma at $2 \pi f \sim T_0$, see~\cite{Ghiglieri:2015nfa, Ghiglieri:2020mhm, Ringwald:2020ist} for more details. The peak frequency of $\Omega_{\rm GW, 0}(f)$ in Eq.~\eqref{eq:CGMB} is in the $(1-100) \, \rm GHz$ range today and depends on the number of entropic relativistic degrees of freedom $g_{*s}(T = T_{\rm max})$.
The peak of $\Omega_{\rm GW, 0}(f)$ approaches the BBN bound if $T_\text{max} \sim M_p$. The CMB constraints on the tensor-to-scalar ratio however constrain the maximal reheating temperature to $T_\text{max} < 10^{16}$~GeV~\cite{Akrami:2018odb} under the assumption of slow-roll inflation and instantaneous reheating. Therefore the detection of the cosmic gravitational microwave background corresponding to $T_{\rm max} > 10^{16} \, \rm{GeV}$ would rule out slow-roll inflation as a viable pre hot Big Bang scenario. Note that since at leading order $\Omega_{\rm GW,0}(f)$ scales linearly with $T_{\rm max}$ and the peak frequency depends on $g_{*s}(T_{\rm max})$, the detection of the peak of the cosmic gravitational microwave background would determine both $T_{\rm max}$ and $g_{*s}(T_{\rm max})$, see~\cite{Ringwald:2020ist} for more details.


\subsubsection{Phase transitions}
\label{sec:PhaseTransitions}

A first order phase transition in the early Universe proceeds by the nucleation of bubbles of the low-temperature phase as the Universe cools below the critical temperature~\cite{Steinhardt:1981ct,Hogan:1984hx}.  Due to the higher pressure inside, the bubbles expand and collide, and the stable phase takes over.  The process disturbs the fluid, generating shear stresses and hence GWs~\cite{Witten:1984rs,Hogan:1986qda}. As the perturbations are mostly compression waves, they can be described as sound waves, and
their collisions are the main source of GWs~\cite{Hindmarsh:2013xza,Hindmarsh:2015qta,Hindmarsh:2017gnf}. \par
The peak frequency of an acoustic contribution to a relic GW background from a strong first order transition is controlled by the temperature of the transition $T_{\ast}$, and the mean bubble separation $R_*$.\footnote{The subscript $*$ means that the corresponding quantity is evaluated at the bubble nucleation time.} Numerical simulations show for wall speeds not too close to the speed of sound that~\cite{Hindmarsh:2017gnf}
\begin{equation}
    f \simeq 26 \left( \frac{1}{H _\ast R _\ast} \right)
\left(  \frac{T_{\ast} }{10^5 \, \text{GeV}} \right) \left( \frac{g_\ast (T_\ast)}{100 } \right)^{1/6}   \text{mHz} \,,
\label{eq:PTf}
\end{equation}
where $H_\ast$ is the Hubble rate at nucleation. The theoretical expectation is that $1 \lesssim (H _\ast R _\ast )^{-1} \lesssim 10^4$.
The intensity depends on $H_\ast R_\ast$, on the fraction of the energy density of the Universe which is converted into kinetic energy $K$ and on the lifetime of the source, which can last for up to a Hubble time.
Denoting the lifetime of the velocity perturbations by $\tau _{\rm v}$, the peak GW amplitude can be estimated as \cite{Hindmarsh:2015qta,Guo:2020grp}
\begin{equation}
\Omega_{\text{GW},0} \simeq 3 \,(H _\ast  R _\ast) \, \left(1- \frac{1}{\sqrt{1+2 H_\ast \tau_{\rm v}}}\right) \left( \frac{100}{g_\ast (T_\ast)} \right)^{1/3} \, K^2 \, \tilde\Omega_\text{GW} \, \Omega_{\text{rad},0} \,,
\label{eq:PTOmega}
\end{equation}
where $\tilde\Omega_\text{GW} $ is a simulation factor and $\tau _{\rm v} = R_\ast /\sqrt{K}$ is the life time of the sound waves. Numerical simulations indicate $\tilde\Omega_\text{GW} = \mathcal{O}\left(10^{-2}\right).$
Hence, $\Omega_{\text{GW},0} \lesssim 10^{-7}$ today, with the upper bound reached only if most of the energy available in the phase transition
is turned into kinetic energy.  This is only possible if there is significant supercooling. \par
The calculation of the kinetic energy fraction and the mean bubble separation requires a knowledge of the free energy density $f(T,\phi)$, a function of the temperature and the scalar field (or fields) $\phi$ whose expectation value determines the phase. If the underlying quantum theory is weakly coupled, and the scalar particle corresponding to $\phi$ is light compared to the masses gained by gauge bosons in the phase transition, this is easily calculated, and shows that first order transitions are generic in gauge theories in this limit \cite{Kirzhnits:1972iw,Kirzhnits:1976ts}, meaning that there is a temperature range in which there are two minima of the free energy as a function of $\phi$.  The critical temperature is defined as the temperature at which the two minima are degenerate, separated by a local maximum. \par

The key parameters to be extracted from the underlying theory, besides the critical temperature $T_c$, are the nucleation rate $\beta$, the strength parameter $\alpha$ and the bubble wall speed $v_w$. The nucleation rate parameter $\beta = d \log p/dt$, where $p$ is the bubble nucleation rate per unit volume, is calculable from $f(T,\phi)$ through an application of homogeneous nucleation theory \cite{Langer:1969bc} to high-temperature fields \cite{Linde:1981zj}.
This calculation also gives $T_\ast $ as the temperature at which the volume-averaged bubble nucleation rate peaks.
The strength parameter is roughly, but not precisely, one quarter of the latent heat divided by the thermal energy (see~\cite{Hindmarsh:2019phv} for a more precise definition) at the nucleation temperature, and  also follows from knowing  $f(T,\phi)$.
The wall speed is a non-equilibrium quantity, which cannot be extracted from the free energy alone, and is rather difficult to calculate accurately (see \cite{Dorsch:2018pat, Laurent:2020gpg} and references therein). In terms of these parameters, it can be shown that \cite{Enqvist:1991xw} $ R_\ast \sim  v_w/\beta  \,.$
The kinetic energy fraction $K$, can be estimated from the self-similar hydrodynamic flow set up around an isolated expanding bubble, whose solution can be found as a function of the latent heat and bubble wall velocity by  a simple one-dimensional integration~\cite{Turner:1992tz,Espinosa:2010hh,Hindmarsh:2019phv}\footnote{Approximate fits can be found in \cite{Espinosa:2010hh}.} and typically ranges between $K = 1-10^{-6}$. \par

Current projected sensitivities for the Einstein Telescope and the Cosmic Explorer can probe a cosmological first order transition occurring at a temperature that is at most a few hundred TeV assuming a modest amount of supercooling \cite{Evans:2016mbw,Punturo:2010zz,Hild:2010id} (i.e.\ when $T_\ast \sim T_C$ and $(R_\ast H_\ast )^{-1} \gtrsim 100$). Recently there has been much interest in high scale transitions motivated by U(1)$_{B-L}$ breaking for leptogenesis and the seesaw scenario \cite{Jinno:2016knw,Marzo:2018nov,Brdar:2018num,Okada:2018xdh,Hasegawa:2019amx,Okada:2020vvb}, as well as multi-step grand unification breaking patterns such as a Pati-Salam model \cite{Croon:2018kqn,Greljo:2019xan,Brdar:2019fur,Huang:2020bbe}. However, in both cases it is more natural to motivate significantly higher scale transitions and to most naturally probe first order transitions at these scales one requires a detector sensitive to frequencies in the range $f \simeq (10^{-3} - 10^3)$ GHz. Finally, it has been shown that zero-temperature phase transitions are likely to occurr in string models that include warped throats~\cite{GarciaGarcia:2016xgv}. These lead to a GW spectrum whose peak falls in the high-frequency range for small values of the compactification volume and large values of the gravitational warp factor associated with the throat in which the phase transition occurs. \par


\subsubsection{Topological defects }
\label{sec:TopologicalDefects}

Cosmic strings are one-dimensional topological defect solutions which may have formed after a phase transitions in the early Universe, if the first homotopy group of the vacuum manifold associated with the symmetry breaking is non-trivial~\cite{Kibble:1976sj,Jeannerot:2003qv}. They can also be fundamental strings from string theory, formed for instance at the end of brane inflation~\cite{Dvali:2003zj,Copeland:2003bj}, and stretched to cosmological scales. The energy per unit length of a string is $\mu \sim \eta^2$, with $\eta$ the characteristic energy scale (in the case of topological strings, it is the energy scale of the phase transition). Typically, the tension of the strings is characterized by the dimensionless combination $G\mu \sim (\eta/M_p)^2$, e.g.~the current upper bound from the CMB is $G\mu \lesssim 10^{-7}$, whereas GW searches in pulsar timing arrays constrain the tension to $G\mu \lesssim 10^{-11}$. Cosmic strings are energetic objects that move at relativistic speeds. The combination of these two factors immediately suggests that strings should be a powerful source of GWs.

Whenever cosmic strings are formed in the early Universe, their dynamics drive them rather rapidly into an attractor solution, characterized by their energy density maintaining always a fixed fraction of the background energy density of the Universe. This is known as the `scaling' regime. During this regime, strings will collide, possibly exchanging `partners' and reconnecting afterwards. This is known as `intercommutation'. For topological strings the intercommutation probability is $\mathcal{P}=1$, whereas $\mathcal{P}<1$ is characteristic in cosmic superstrings networks. Closed string configurations -- loops -- are consequently formed when a string self-intersects, or two strings cross. Loops smaller than the horizon decouple from the string network and oscillate under their own tension, which results in the emission of gravitational radiation (eventually leading to the decay of the loop). The relativistic nature of strings typically leads to the formation of {\it cusps}, corresponding to points where the string momentarily moves at the speed of light~\cite{Turok:1984cn}. Furthermore, the intersections of strings generates discontinuities on their tangent vector known as {\it kinks}. All loops are typically expected to contain cusps and kinks, both of which generate GW bursts~\cite{Damour:2000wa,Damour:2001bk}. Hence, a network of cosmic (super-)strings formed in the early Universe is expected to radiate GWs throughout the entire cosmological history, producing a stochastic background of GWs from the superposition of many uncorrelated bursts.\footnote{An alternative strategy to the detection of cosmic string networks is the search for sufficiently strong GW transient signals~\cite{Aasi:2013vna,Abbott:2017mem} which do not form part of the stochastic background of GWs.}

A network of cosmic strings contains therefore, at every moment of its evolution (once in scaling), sub-horizon loops, and long strings that stretch across a Hubble volume. The latter are either infinite strings or in the form of super-horizon loops, and are also expected to emit GWs. However, the dominant contribution is generically that produced by the superposition of radiation from many sub-horizon loops along each line of sight.

The power emitted into gravitational radiation by an isolated loop of length $l$ can be calculated using the standard formulae in the weak gravity regime~\cite{Weinberg:1972kfs}. More explicitly, we can assume that, on average, the total power emitted by a loop is given by $P_{\rm 1Loop} = \Gamma \times (\mathrm{G} \mu) \times \mu$, with $\mathrm{\Gamma}$ is a dimensionless constant (independent of the size and shape of the loops). Estimates from simple loops~\cite{Vachaspati:1984gt,Burden:1985md,Garfinkle:1987yw}, as well as results from Nambu-Goto simulations~\cite{Blanco-Pillado:2017oxo}, suggest that $\mathrm{\Gamma}=50$. The GW radiation is only emitted at discrete frequencies by each loop, $\omega_n = 2\pi n/T$, where $T=l/2$ is the period of the loop, and $n=1,2,3,\ldots$ We can write $P_{\rm 1Loop} = \mathrm{G} \mu^2\sum_n P_n$, with $P_n$ characterizing the power emitted at each frequency $\omega_n$ for a particular loop, depending on whether the loop contains cusps, kinks, and kink-kink collisions~\cite{Burden:1985md,Allen:1991bk}. Namely, it can be shown that for large $n$,
$P_{ n}=\frac{\mathrm{\Gamma}}{\zeta(q)}n^{-q}$, where $\zeta(q)$ is the Riemann zeta function. The latter is simply introduced as a normalization factor, to enforce the total power of the loop to be equal to $\mathrm{\Gamma}=\sum_n P_n$. The parameter $q$ takes the values $4/3$,
$5/3$, or $2$ depending on whether the emission is dominated by cusps, kinks or kink-kink collision respectively, see e.g.~\cite{Vachaspati:1984gt,Binetruy:2009vt,Auclair:2019wcv}.

The resulting GW background from the stochastic background emitted by the loops chopped-off along the radiation domination period is characterized by a scale-invariant energy density spectrum, spanning over many frequency decades. The high-frequency cut-off of this plateau is determined by the temperature of the thermal bath at formation of the string network, with $T_\text{max} \lesssim 10^{16}$~GeV implying a turn-over frequency of $f_\Delta \lesssim 10^9$~GeV~\cite{Gouttenoire:2019kij}.
The amplitude of this plateau is given by~\cite{Auclair:2019wcv}
\begin{equation}\label{eq:plateauStringsNG}
\mathrm{\Omega}^{\rm plateau}_{\text{GW},0}(f) \approx 8.04\, \Omega_{\text{rad},0} \sqrt{\frac{G\mu}{\mathrm{\Gamma}}}\,.
\end{equation}
In particular, this does not depend on the exact form of the loop's power spectrum, nor on whether the GW emission is dominated by cusps or kinks, but rather depends only on the total GW radiation emitted by the loops. The amplitude in Eq.~\eqref{eq:plateauStringsNG} indicates that the stochastic GW background from cosmic strings can be rather large.\footnote{{\it Important remark}: as the characteristic width $\delta \sim 1/\eta$ of a cosmic string is generally much smaller than the horizon scale, it is commonly assumed that strings can be described by the Nambu-Goto (NG) action, which is the leading-order approximation when the curvature scale of the strings is much larger than their thickness. The above plateau amplitude Eq.~\eqref{eq:plateauStringsNG} applies only for the case of NG strings. For NG strings to reach scaling, the GW emission of the loops is actually crucial, as it is the loss of loops from the network that guarantees the scaling. The loops need therefore to decay in some way (this is precisely what the GW emission takes care of), so that their energy is not accounted anymore as part of the string network. However, in field theory simulations of string networks~\cite{Vincent:1997cx,Hindmarsh:2008dw,Daverio:2015nva,Hindmarsh:2017qff}, the network of infinite strings reaches a scaling regime thanks to energy loss into classical radiation of the fields involved in the simulations. The simulations show the presence of extensive massive radiation being emitted, and that the loops formed decay within a Hubble time. This intriguing discrepancy has been under debate for the last $\sim$ 20 years, but the origin of this massive radiation in the lattice simulations is not understood.}

Moreover, if the phase transition responsible for cosmic string formation is embedded into a larger grand unified group, then, depending on the structure of that larger group, cosmic strings may be only metastable, decaying via the (exponentially suppressed) production of monopoles~\cite{Vilenkin:1982hm,Monin:2008mp,Monin:2009ch,Leblond:2009fq}. In this case, the low-frequency end of the spectrum, corresponding to GW emission at later times, will be suppressed and the signal may \textit{only} be detectable in the high-frequency range~\cite{Leblond:2009fq,Dror:2019syi,Buchmuller:2019gfy,Buchmuller:2020lbh}. In this case, the string tension is only bounded by the BBN bound, $G \mu \lesssim 10^{-4}$ and the scale-invariant part of the spectrum may extend from $10^3$~Hz (LIGO constraint) up to $10^9$~Hz (network formation).

Finally, let us recall that long strings (infinite and super-horizon loops) also radiate GWs. One contribution to this signal is given by the GWs emitted around the horizon scale at each moment of cosmic history, as the network energy-momentum tensor adapts itself to maintaining scaling~\cite{Krauss:1991qu,JonesSmith:2007ne,Fenu:2009qf,Figueroa:2012kw,Figueroa:2020lvo}. This background is actually expected to be emitted by any network of cosmic defects in scaling, independently of the topology and origin of the defects~\cite{Figueroa:2012kw}. It represents therefore an irreducible background generated by any type of defect network. In the case of cosmic string networks modelled by the Nambu-Goto approximation (where the thickness of the string is taken to be zero), this background represents a very sub-dominant signal compared to the GW background emitted from the loops. In the case of field theory strings (for which simulations to date indicate the absence of `stable' loops), it is instead the only GW signal (and hence the dominant one) emitted by the network.

The GW energy density spectrum of this irreducible background from long strings is predicted to be exactly scale-invariant, for the modes emitted during radiation domination~\cite{Figueroa:2012kw}. The power spectrum of this background therefore mimics the spectral shape of the dominant signal from the loop decay, but with a smaller amplitude. The amplitude of the irreducible GW background from string networks depends ultimately on the fine details of the so called \textit{unequal-time-correlator} of the network's energy-momentum tensor. This correlator, however, can be obtained only  accurately from sufficiently large scale lattice simulations. In the case of global defects, the scale-invariant GW power spectrum has been obtained numerically with massively parallelized lattice field theory simulations for global strings\footnote{The irreducible background from the more interesting case of Abelian-Higgs lattice field theory simulations, has unfortunately not been yet studied.} as~\cite{Figueroa:2012kw}
\begin{equation}\label{eq:GWSOSFplateau}
    \mathrm{\Omega}_{\text{GW}, 0}\simeq 4\times 10^4\, \mathrm{\Omega}_{\text{rad},0} (G\mu)^2\,.
\end{equation}
Despite the numerical prefactor being much larger than unity, the quadratic scaling proportional to $(G\mu)^2$ suppresses significantly the amplitude, see e.g.~\cite{Buchmuller:2013lra} for a comparison among GW signals emitted from the same string network. This amplitude is clearly subdominant when compared to the amplitude of the GW signal from the loops, which scales as $(G\mu)^{1/2}$,  see Eq.~\eqref{eq:plateauStringsNG}. A proper assessment of the power spectrum of this stochastic background requires further results not available yet; namely, lattice simulations of cosmic networks with a larger dynamical range.

Finally, we point out that since the irreducible GW emission described before is expected from any scaling defect network, global texture networks also emit a GW background due to their self-ordering during scaling~\cite{JonesSmith:2007ne,Fenu:2009qf,Giblin:2011yh,Figueroa:2012kw, Figueroa:2020lvo}. Textures are formed when the second (or higher) homotopy group of the vacuum manifold is non-trivial~\cite{Vilenkin:2000jqa}. One can achieve such condition in either case of the symmetry breaking of a global or a gauge group. In the case of a global group, the GW spectrum is scale invariant for radiation domination~\cite{JonesSmith:2007ne,Fenu:2009qf,Figueroa:2012kw}, and exhibits a peak at the horizon today for matter domination~\cite{Figueroa:2020lvo}. For a gauge group the GW spectrum is suppressed at low-frequencies as the massive gauge boson can prevent self-ordering as the gauge field can cancel the gradient field at large scales. The peak frequency and amplitude of a gauge texture is therefore set by the symmetry breaking scale $v$~\cite{Dror:2019syi},
\begin{equation}
    f \sim  \left(\frac{v}{M_p}\right) \, 10^{11} \, {\rm Hz} \,, \qquad
    \Omega_{\text{GW},0} \sim 2 \times 10^{-4} \, \left(\frac{v}{M_p}\right)^4 \,.
    \label{eq:gaugetextures}
\end{equation}
Given the frequency and amplitude both increase with $v$, it is only reasonable to consider high-frequency signals.

\subsubsection{Evaporating primordial black holes}
\label{sec:EvaporatingPBHs}

Sec.~\ref{sec:PBHmergers} discussed the GW signal emitted by primordial BHs that merge in the late Universe. Note that light primordial BHs (with mass smaller than $10^{11} \, {\rm kg}$) that evaporate before BBN could produce a stochastic spectrum of GWs by merging and scattering~\cite{Dolgov:2011cq}. The typical frequency for such a source is at $f \gtrsim \rm{GHz}$. Here we consider yet another source of GWs tied to primordial BHs, namely the emission of gravitons as part of the Hawking radiation of primordial BHs. This is particularly relevant for light primordial BHs which evaporated before BBN  and for primordial BHs which have life time ranging from BBN until today ($10^{11} \, {\rm kg} \lesssim m_{\rm PBH} \lesssim 10^{14} \, {\rm kg}$).

The graviton emission from a population of primordial BHs induces a stochastic background of GWs~\cite{Anantua:2008am,Dong:2015yjs} that peaks at very high-frequencies, between $f\sim 10^{13} \, {\rm Hz}$ and $ 10^{22} \, {\rm Hz}$.
The shape and amplitude of the GW frequency spectrum depends on multiple factors, such as primordial BH abundance at formation, their mass spectrum, their eventual spin, the number of degrees of freedom in the particle theory. Roughly speaking, due to the redshift of GW amplitude and frequency, the observed GW spectrum is dominated by the latest stages of the primordial BH evolution and the frequency is hence set by the evaporation time (and hence the initial mass) of the primordial BH.

Taking into account the limits on the primordial BH abundance from BBN and extra-galactic background radiation, the maximum amplitude can be up to $\Omega_{\text{GW},0} \approx 10^{-7.5}$ for primordial BHs evaporating just before BBN, $m_{\rm PBH} \lesssim 10^{11} \, {\rm kg}$.  For heavier ones that might have not yet totally evaporated today, $10^{11} \, {\rm kg} \lesssim m_{\rm PBH} \lesssim 10^{14} \, {\rm kg}$, it can be up to $\Omega_{\text{GW},0} \approx 10^{-6.5}$~\cite{Dong:2015yjs}, with a spectrum peaked at frequencies between $10^{18}$ Hz and $10^{22}$ Hz.

The most interesting case are possibly much lighter primordial BHs which would have completely evaporated well before BBN, e.g.\ if they are produced at an energy close to the grand unification scale, $E \sim 10^{15} \, {\rm GeV}$. Because the primordial BH density decreases like $\propto 1/a^3$ and the radiation density is $\propto 1/a^4$, such early decaying primordial BHs can be very abundant in the early Universe, leading to an early matter dominated phase. GWs produced in their decay could then constitute a sizable fraction of the subsequent radiation dominated phase, limited only by the BBN and CMB constraints (see Eq.~\eqref{eq:BBNbound}). For primordial BHs from the rand unification scale, the GW frequency spectrum has a peak around $10^{15}$ Hz and can reach an amplitude $\Omega_{\text{GW},0} \sim 10^{-8}$ for number of degrees of freedom $n_{\rm dof}\sim 10^3$~\cite{Anantua:2008am}.  It might therefore be in the range of detection of future instruments.

\subsection{Miscellaneous}
\label{sec:Misc}

In this section we summarize a few more ideas that have been shown to be relevant for high-frequency GW production but require more exotic setups to be realized and do not find place in the classification proposed above.

\begin{itemize}
    \item \textit{Brane-world scenarios}: the brane-world scenario~\cite{Rubakov:1983bb} proposes that the very weak force of gravity in our $(3+1)$-dimensional Universe is only a part of the full strength gravity which is felt in the fifth dimension at a level commensurate with the other forces. This scenario suggests that two $(3+1)$-dimensional branes - one of which represents our $4$-dimensional Universe - are separated in a fifth dimension by a small distance~\cite{Randall:1998uk, Maartens:2010ar}. If violent gravitational events - such as BH mergers - take place on the `shadow' brane, which is in close proximity to our brane - they would excite oscillations not only in the shadow brane but also in the fifth dimensional brane separation, leading to GW production on our visible brane~\cite{Seahra1, Seahra2}.
    \item \textit{Pre Big Bang cosmology}: the pre Big Bang scenario provides an alternative to cosmological inflation to provide the initial conditions for the hot Big Bang theory. The scenario exploits the fundamental symmetries of string theory to build a model in which the Universe starts in a cold and empty state in the infinite past and moves towards a state of high curvature through an accelerated expansion~\cite{Gasperini:2002bn, Gasperini:2007vw}. The state of high curvature corresponds to a region in the parameter space in which the theory is strongly coupled. It is then assumed that the strongly coupled theory is able to match this initial accelerated expansion to the usual hot Big Bang cosmology. Interestingly, this scenario predicts a blue spectrum of GWs, with a peak at high-frequency~\cite{Brustein:1995ki}.
    \item \textit{Quintessential inflation}: if the inflationary epoch is followed by a phase in which the equation of state is stiffer than radiation, the stochastic spectrum of GWs features a growth at high-frequency, followed by a sharp cutoff~\cite{Giovannini:1999bh}. This kind of behaviour is expected in quintessential models of inflation, such as the one investigated in~\cite{Peebles:1998qn}. The position of the peak depends very weakly on the number of minimally coupled scalar fields of the model, but it is independent of the final curvature at the end of inflation. Therefore, it is located at $100 \, \text{GHz}$ and cannot be moved around. The amplitude of the GW spectrum can become very large: in~\cite{Giovannini:1999bh} the authors present a choice of the parameters such that $\Omega_{\rm GW, 0} \simeq 10^{-6}$ at the peak.
    \item \textit{Magnetars}: magnetars are neutron stars that feature very large surface magnetic fields $\sim 10^9-10^{11} \, \text{T}$. Ref.~\cite{Wen:2017itr} suggests that gamma-ray bursts produced by the magnetar itself or by a companion body forming a binary system, and interacting with the surface magnetic field of the magnetar could be the source of high-frequency GW, with frequency around $10^{20} \, \text{Hz}$ and energy density up to $\Omega_{\rm GW, 0} \sim 10^{-6}$.
    \item \textit{Reheating}: Ref.~\cite{Ema:2020ggo} shows that there exists a model-independent contribution to the stochastic GW background, due to the oscillations of the inflaton (or any other scalar field that oscillates around its minimum after inflation) around the minimum of its potential during preheating. These oscillations provide a driving force in the equation of motion for the tensor modes, leading to GW production at high-frequency $\gtrsim 10^5 \, \text{Hz}$. The amplitude of this signal is bound to be quite small: in Ref.~\cite{Ema:2020ggo} the authors present a choice of parameters such that $\Omega_{\rm GW, 0} \lesssim 10^{-21}$.
    \item \textit{Thermal gravitational noise of  the Sun}: the thermal motion of charged particles in the plasma of protons and electrons present at the core of stars leads to the production of a gravitational radiation noise~\cite{Weinberg:1972kfs, BisnovatyiKogan:2004bk}. The frequency of this radiation is determined by the frequency of the collisions and, in the case of the Sun, it falls in the range $(10^{12}-10^{18})\,$ Hz.
    \item \textit{Plasma instabilities}: in Ref.~\cite{Servin:2003cf} the authors studied the interaction of electromagnetic waves and GWs in a magnetised plasma. In the high-frequency regime, a circularly polarized electromagnetic wave travelling parallel to the background magnetic field present in a plasma generates a GWs with the same frequency of the electromagnetic wave.
\end{itemize}

In summary, a broad range of theoretically well motivated extensions of the Standard Model of particle physics predict the existence of GW sources at different stages in the evolution of our Universe. The corresponding GW frequency range and amplitudes are summarized in Fig.~\ref{fig:hc_stochastic_summary} and in Fig.~\ref{fig:hc_coherent_summary} as well as in Tab.~\ref{tab:summary-stochastic} and Tab.~\ref{tab:summary-coherent} in App.~\ref{sec:SummaryTable}.


\section{Detection of gravitational waves at high-frequencies}
\label{sec:exp}

After the first detection of GWs at frequencies in the range $(0.1 - 2.0)\,$ kHz~\cite{PhysRevX.9.031040}, the expansion into other frequency bands is a natural next step -- as it was in the 1950's when radio, X-Ray and UV observations became possible with new technology. As detailed in the previous section, many exciting questions in astronomy, cosmology and fundamental physics are tied to GW signals with frequencies (far) above the capabilities of current detectors or their upgrades. Fig.~\ref{fig:hc_stochastic_summary} and Fig.~\ref{fig:hc_coherent_summary} give an impression of the range of GW amplitudes expected for various coherent and stochastic sources.
Even GW upper limits with no known source targets at the time of publication {of this white paper} may be valuable in restricting physical theories.

In this section, we will investigate the experimental possibilities for the detection of high-frequency GWs. First, we will give an overview of current GW detectors and their limitations, followed by the introduction of several concepts for the detection of high-frequency signals.
{Depending on the detector concept and the targeted sources, we quote detector sensitivities in terms of strain $\sqrt{S_n}$ or in terms of the dimensionless quantities $h_{c,n}$ (for stochastic signals) and $h_{0,n,mono}$ (for monochromatic signals), see Sec.~\ref{sec:Notation} for details. Careful consideration of operation and bandwidth is needed to convert between these quantities. }

\subsection{Laser interferometers and resonant mass detectors and their limitations}
\label{sec:detectors}

The first GWs were detected by the Advanced LIGO~\cite{PhysRevLett.116.131103} detectors in the US and the Advanced Virgo detector in Italy~\cite{AdvVirgo}. In early 2020, the Japanese KAGRA detector~\cite{PhysRevD.88.043007} joined LIGO's third observing run. These detectors are all based on the principle of a Michelson interferometer, using large suspended mirrors with several kilometers distance between them. Several other detectors are in the design phase. These  detectors typically have their peak sensitivity at frequencies of a few hundred Hz.

However, some future detectors are designed to particularly expand the detection band towards either low or high-frequencies.
To expand the detection band of Earth-bound interferometers to frequencies below 10\,Hz, cryogenically cooled mirrors, large beam diameters and operation underground are considered~\cite{ETdesign,Adhikari_2020}.
LISA, also based on laser interferometry, is a planned, satellite-based detector to increase the arm length beyond the possibilities on Earth and to reduce environmental noise sources such as seismic noise~\cite{Audley:2017drz}. LISA will have its peak-sensitivity in the mHz range. To increase interferometer sensitivity towards higher frequencies, options are an increase of laser power and/or resonant operation. The planned Australian NEMO detector will be targeting frequencies of up to several kHz, see Sec.~\ref{sec:ozgrav} below.\footnote{We note that through the GW memory effect~\cite{Christodoulou:1991cr,Thorne:1992sdb}, these interferometers are sensitive to high-frequency GW bursts far beyond their nominal frequency band~\cite{mcneill2017gravitational,ebersold2020search}. }

While increasing the arm-length of an interferometer increases strain signal in some frequency band, longer arms are only really beneficial as long as the GW wavelength is longer than the interferometer arms. For significantly higher frequencies (MHz) interferometers with arm-lengths of meters are more suitable, but are of course at the same time limited by the smaller strain sensitivity achievable with shorter arms. This constitutes the main limitation of laser interferometers, used as direct strain meters,
towards higher GW signal frequencies.

A concept to detect GWs which existed prior to the interferometers are resonant bar detectors, initially proposed and built by Joseph Weber in the 1960s. Their modern successors, resonant spheres, have peak sensitivities at several kHz. In Sec.~\ref{sec:spheres}, we will give a summary of these resonant spheres.

\subsubsection{Laser Interferometers: Neutron Star Extreme Matter Observatory (NEMO)} \label{sec:ozgrav}

The first detection of a binary neutron star merger coalescence in 2017~\cite{abbott2017gw170817} has increased the interest in the development of GW detectors with sensitivity in the few kHz regime which will be capable of detecting the merger and ringdown part of the waveform~\cite{martynov2019exploring}. It is expected that such detectors will need to have strain sensitivities approaching $\sqrt{S_n} \simeq 10^{-24}\, \text{Hz}^{-1/2}$ in the range $(1-4)\,$kHz for events that are likely to occur a few times per year. This sensitivity should be achieved by the third generation terrestrial GW detectors that are anticipated to come online in the later half of the 2030s~\cite{Evans:2016mbw,Punturo:2010zz}. The Australian GW community is currently exploring the feasibility of a new detector, `NEMO', dedicated to detecting this merger phase and the following ringdown as well as testing third generation technology on a smaller scale~\cite{ozhf, Bailes:2019oma,Adya2020}. The planned sensitivity of this detector would reach $\sqrt{S_n} \simeq 10^{-24}\, \text{Hz}^{-1/2}$ in the range $(1 - 2.5)\,$kHz~\cite{ozhf}. This detector will work in collaboration with the existing second generation GW detector network that will provide sky localization for electromagnetic follow-up.

The dominant high-frequency noise source for interferometric GW detectors is quantum phase noise or shot noise as it is otherwise called. The magnitude of this noise source is inversely proportional to the square of the product of the circulating power incident on the test masses and the length of the arms of the detector. This generally necessitates extremely high powers in the arms of the interferometers ($\approx 5 \,\text{MW}$ in the case of NEMO). Such high circulating powers lead to technical issues such as parametric and tilt instabilities and thermal induced distortions. These issues can be challenging to deal with, however a dedicated high-frequency detector promises to make this easier. This is because low-frequency sensitivity limits the actuation that can be applied to the test masses to correct instabilities and distortions. Further, relaxing the low-frequency sensitivity relaxes the requirements on seismic isolation and test mass suspension systems that can significantly reduce the cost.

\subsubsection{Interferometers up to 100\,MHz}
\label{sec:100MHzInterferometers}

As was first pointed out by Mizuno~\cite{Mizuno}, in laser interferometers the overall stored energy in the form of circulating laser power sets a limit on the achievable sensitivity and bandwidth, which is a consequence of the quantum Cram\'{e}r-Rao bound. For a given laser power, higher bandwidth needs to be traded in for an increase in sensitivity. While opto-mechanical resonances can be introduced in the signal response of interferometers to shape the sensitivity curve for specific frequencies~\cite{Somiya:2016pla,Korobko:2018pla}, it appears unlikely that the stored laser power can be further increased by several orders of magnitude. Therefore, broadband interferometric detectors reaching into the MHz detection range (while maintaining LIGO or Virgo-level strain sensitivity) seem not to be a viable option when taking also the arm-length argument from above into account.

Nevertheless there are three notable efforts (two existing and one under construction) of laser interferometers in the MHz range, which currently set the best experimental upper limits on GWs in their respective frequency bands.

One option is to build kHz-bandwidth interferometric detectors that are centered around much higher frequencies. Akutsu et al.~\cite{akutsu} have published upper limits from such a system working at 100 MHz. The detector used a synchronous recycling architecture based on a resonant recycling cavity of dimension 75\,cm and a Nd:YAG laser with a power output of 0.5 W. The limit on stochastic GW signals was reported to be $\sqrt{S_n} \sim 10^{-16}~\text{Hz}^{-1/2}$, placing a bound of $h_{c,\rm sto} \lesssim 7 \times 10^{-14}$. A study of the potential of this technique~\cite{Nishizawa} showed that a sensitivity of $10^{-20}~\text{Hz}^{-1/2}$ is possible at 100\,MHz with a bandwidth of 2\,kHz, but the sensitivity decreases with increasing frequency and is not competitive above 1\,GHz.

The sensitivity of a single instrument can be surpassed by correlating two co-located instruments in the case of searching for stochastic signals from GWs or other sources.
The \emph{Holometer} experiment at Fermilab consists of two co-located power recycled Michelson interferometers with 40-meter long arms. While their primary research target has been signatures of quantization of spacetime, they reach a sensitivity of $10^{-21} \, \text{Hz}^{-1/2}$ approximately in the band $(1-13)\,$MHz~\cite{PhysRevD.95.063002}. Using a 704-hr dataset from the Holometer experiment, the authors of Ref.~\cite{Martinez:2020cdh} concluded that there are no identifiable harmonic sources such as cosmic string loops and eccentric BH binaries emitting in the frequency range $(1-25) \, \text{MHz}$.

The experimental GW group at Cardiff University is planning a set of two wide-band table-top interferometers sensitive in the band $(1-100)\,$MHz~\cite{vermeulen2020experiment}.
These will be able to set new upper limits on a stochastic GW background in this frequency band.

\subsubsection{Spherical resonant masses}
\label{sec:spheres}

The principle of a \textit{resonant mass detector} is that its vibrational eigenmodes can get excited by a GW. These mechanical oscillations are transformed into electrical signals, using electromechanical transducers, and amplified by electrical amplifiers. These resonant detectors have a relatively small bandwidth, usually of less than 100~Hz. Thermal noise, Johnson-Nyquist noise, pump phase noise (if the transducer is parametric), back-action noise, and amplifier noise are the internal noises of this kind of detector. Therefore, the resonant mass antenna and transducers are made of high-quality factor
materials in order to decrease thermal (mechanical) and Johnson-Nyquist (electrical) noises.

The idea of a spherical resonant mass antenna for GW detection has a long history and was first proposed by Robert L. Forward in 1971~\cite{Forward1971} followed by several decades of exploration and proposals~\cite{wagoner1977multimode, hamilton1990resonant, PhysRevLett.70.2367}.
In 1991, Aguiar  proposed a large spherical antenna project in Brazil~\cite{Aguiar:2010kn}. This detector, Mario Schenberg, in S\~ao Paulo, Brazil\cite{Da_Silva_Costa_2014}, was started to be built in 2000, around the same time as Mini-GRAIL, in Leiden, Netherlands. These two spherical detectors were active for about 15 years. At present, they are decommissioned, but Schenberg is planned to be reassembled at INPE, in S\~ao Jos\'e dos Campos, about 100 km from its initial site at the University of S\~ao Paulo.\footnote{These detectors had much smaller masses and diameters than originally proposed in the ’90s of up to 120\,Tons and 3\,m, respectively, resonant around $\sim 700\,$ Hz.} Such detectors have a bandwidth of 50-100 Hz with peak frequencies around 3 kHz for the quadrupole modes. To increase the frequency range, a xylophone configuration of several spheres has been proposed~\cite{PhysRevD.54.2409}.

Spherical antennas provide more information, compared to the classical bar antennas, because of their quadrupole modes, while also being significantly more sensitive due to their favorable geometry of having a larger cross-section at identical mass. From the output of six 6 transducers tuned to the quadrupole modes of the sphere, a single sphere can obtain complete information about the polarization and direction of the incoming wave.

In 2004, Mini-GRAIL reached a peak strain sensitivity of $\sqrt{S_n} \simeq 1.5 \times 10^{-20} \;\text{Hz}^{-1/2}$ at a frequency of 2942.9\,Hz at temperatures of 5\,K. Over a bandwidth of 30\,Hz, the strain sensitivity was  about $\sqrt{S_n} \simeq 5 \times 10^{-20}\; \text{Hz}^{-1/2}$~\cite{Gottardi:2007zn}.
Schenberg operating also at 5\,K, reached strain sensitivities of $\sqrt{S_n} \simeq 1.1 \times  10^{-19}\; \text{Hz}^{-1/2}$ for its quadrupolar modes ($\sim 3.2 \,$ kHz) and $\sqrt{S_n} \simeq 1.2 \times  10^{-20}\; \text{Hz}^{-1/2}$ for its monopolar mode ($\sim 6.5 \,$ kHz), in 2015. Both antennas could reach sensitivities around $\sqrt{S_n} \simeq 10^{-22}\; \text{Hz}^{-1/2}$ when operating at 15\,mK. Schenberg, because it uses parametric transducers, can reach higher sensitivities if it implements squeezing of the signal. In this case, it would have similar sensitivities as the ultimate sensitivities of Advanced LIGO and Virgo around 3.2\,kHz.\footnote{However, the interferometric detectors have a peak sensitivity around one order of magnitude higher at a few hundred Hz.}

The conceptual difficulties in pushing this technology to higher frequencies are similar to the issues discussed for laser interferometers. Searching for GWs at higher frequencies requires smaller resonating spheres and consequently requires measuring smaller absolute displacements to achieve the same strain sensitivity. Note also that contrary to laser interferometers, resonant mass detectors have not yet reached the standard quantum limit yet. It thus seems unlikely that this technology can be pushed significantly beyond the kHz region.


\subsection{Detection at frequencies beyond current detectors}

In this section we will introduce several ideas and concepts for the detection of GWs at high-frequencies beyond the capabilities of currently existing GW detectors.

\subsubsection{Optically levitated sensors}
\label{sec:OpticallyLevitatedSensors}

\textit{Optically levitated dielectric sensors} have been identified as a promising technique for resonant GW searches spanning a wide frequency band from a few $\sim$ kHz to $\sim 300$ kHz \cite{arvanitaki:2016gw,aggarwal2020searching}. A dielectric nano-particle suspended appropriately at the anti-node of a  laser standing wave within an optical cavity will experience a force when a passing GW causes a time-varying strain of the physical length of the cavity. The particle will be displaced from the location of the trapping light anti-node, resulting in a kick on the particle at the frequency of the GW space-time disturbance. The trapping frequency and mechanical resonance linewidth are widely tunable based on the laser intensity and laser cooling parameters chosen. For possible sources in this frequency band see e.g.~Sec.~\ref{sec:NS}, Sec.~\ref{sec:PBHmergers} and Sec.~\ref{sec:ECOs}.

When detecting the resulting displacement of the particle at the trapping resonance frequency, the sensitivity is limited by Brownian thermal noise in the particle itself rather than the displacement detection of the particle. This results in improved sensitivity at higher frequency (unlike traditional interferometer style detectors which decrease sensitivity at high-frequency due to laser shot noise)~\cite{arvanitaki:2016gw}.  The low-friction environment made possible by optical levitation in ultra-high vacuum enables extremely sensitive force detection~\cite{Ranjit:2016zn}, which becomes ultimately quantum-limited by photon-recoil heating from discrete scattering events of individual trap laser photons~\cite{Jain:2016re}.

A 1-meter prototype Michelson-interferometer configuration detector called the `Levitated Sensor Detector' is under construction at Northwestern University in the US, with a target sensitivity of better than $\sqrt{S_n}  \sim {10}^{-19} \, \text{Hz}^{-1/2}$ at $10$ kHz and $\sqrt{S_n} \sim 10^{-21} \, \text{Hz}^{-1/2}$ at $100$ kHz~\cite{aggarwal2020searching}. With Prof. P. Barker and coworkers at partner institution University College London, fiber-based approaches are being investigated to permit longer cavities without the need for expensive optics~\cite{Pontin:2018fi}. The ultimate strain sensitivity of a $10$-meter room-temperature instrument is estimated to be better than approximately $\sqrt{S_n}  \sim {10}^{-20} \, \text{Hz}^{-1/2}$ at $10$ kHz and $\sqrt{S_n}  \sim 10^{-22} \, \text{Hz}^{-1/2}$ at $100$ kHz. For a cryogenic $100$-meter instrument this can be improved by more than an order of magnitude across much of the frequency band~\cite{aggarwal2020searching}. A detailed analysis of the search reach for GWs produced by axions via the BH superradiance process is provided in Ref.~\cite{aggarwal2020searching}.

\subsubsection{Inverse Gertsenshtein effect}
\label{sec:inv_gertsenshtein}
The \textit{Gertsenshtein effect} describes the conversion of photons to GWs in the presence of a magnetic field and was considered already decades ago as a source of GWs~\cite{Gertsenshtein}. While the coupling constant for this process is too small to be of interest for experiments in the near to medium future, the inverse Gertsenshtein effect, frequently referred to as magnetic conversion, can indeed be used to search for GWs~\cite{Boccaletti1970,Fuzfa:2015oaa,Fuzfa:2017ana}. While such dedicated instruments do not exist yet (apart from small prototypes), a first step in this direction has been done by using existing data from axion-search experiments. In these experiments, typically a strong static magnetic field of several Tesla is set up with field lines perpendicular to some interaction region, through which a beam line passes. In their nominal usage these experiments would search for axion-like particles, which can convert to photons in the presence of the magnetic field. These photons would be detected at the end of the beam line by electromagnetic detectors within the frequency band of interest (e.g. photodetectors for optical radiation).
The very same experimental arrangements can also be used to search for GWs, by re-interpreting the acquired data,
as has been pointed out and performed by the work in Ref.~\cite{Ejlli2019}.
This work could set first upper limits for GWs at optical and X-ray frequencies (i.e.\ around 500\,THz and $10^{6}$\,THz respectively).

In the future this class of experiment is expected to continue with larger detectors of this sort being constructed. Given the motivation in the context of searches for high-frequency GWs, a dual usage of these detectors could be imagined
with dedicated instruments and operational modes to search for GWs. For example, the planned IAXO detector~\cite{Armengaud:2014gea} aims at searching for axions produced in the core of the sun. If it were  fitted with different electromagnetic receivers, from radio to optical frequencies, GW searches could be facilitated in these bands. A particular advantage of searches with IAXO would be that the device can be pointed (within some limits) to different points in the sky. It could be of particular interest to point to patches in the sky where, for example, a binary BH merger is predicted to happen. The latter is a real prospect once the LISA space interferometer will be operational, which can detect inspiralling BHs long before their merger.

Note that, in principle, the inverse Gertsenshtein effect might be exploited at all frequencies, and that one further advantage of this concept is the tunability of the solid angle, which changes according to the direction of the magnetic field, see Sec.~\ref{subsec:switch}. In particular, the inverse Gertsenshtein effect has substantial room for development especially at GHz frequencies where many of the early Universe signals converge. Using magnets developed for particle accelerators a conversion path length of 100 metres with uniform magnetic field of 5.6\,T is quite realistic, and being implemented for the ALPS experiment searching for axion-like particles~\cite{Bahre:2013ywa}. Electromagnetic detectors having a thermal noise equivalent to 0.1 Kelvin would lead to a sensitivity around $h_{c,n} \simeq 10^{-26}$. This sensitivity could be further enhanced by the inclusion of a Fabry Perot cavity in the conversion volume and a factor of 100 improvement might be possible. Unfortunately, this improved sensitivity would be gained at the expense of the wide bandwidth of the technique and this would limit the applicability to stochastic signals. Reference~\cite{Ringwald:2020ist} estimates the sensitivity that can be reached in the GHz region by using Single Photon Detectors (SPD) and Heterodyne radio receivers (HET). The corresponding limits are reported in Fig.~\ref{fig:hc_coherent_summary} and Fig.~\ref{fig:hc_stochastic_summary}.

The conversion of GWs to photons by the inverse Gertsenshtein effect cannot only be exploited in a laboratory setting but also by considering astrophysical or even cosmological `detectors', see e.g.~\cite{Pshirkov:2009sf, PhysRevLett.74.634, Dolgov:2012be, Cillis:1996qy,Domcke:2020yzq}. In this case, the magnetic field is weaker and the background is much harder to control, but cosmic magnetic fields can extend coherently over kpc or even Mpc, implying an enormous `detector volume'. The frequency range of MHz to GHz coincides with the Rayleigh-Jeans tail of the cosmic microwave background, a target of existing and upcoming radio telescopes. For example, the data of ARCADE 2~\cite{Fixsen_2011} and EDGES~\cite{Bowman:2018yin} can be recast to respectively constrain $h_{c,\rm sto} < 10^{-24}  (10^{-14})$ in the range $3 \, {\rm GHz} \lesssim f  \lesssim 30 \, {\rm GHz}$ and $h_{c, \rm sto}(f \approx 78 \, {\rm GHz}) < 10^{-12} (10^{-21})$ for the strongest (weakest) cosmic magnetic fields in accordance with current astrophysical data~\cite{Domcke:2020yzq}.

\subsubsection{GW to electromagnetic wave conversion in a static electric field}
\label{sec:ConversionEfield}

Lupanov \cite{Lupanov} considered the inverse Gertsenshtein effect but using a static electric field rather than a static magnetic field. The physics is essentially the same in the two cases but since the intensity of electric fields in laboratory settings is limited by the tendency to pull electrons from nearby conductors (or dielectrics) and thereby cause local short circuits, the available energy densities in electric fields are about one millionth of that created by magnetic fields in the several Tesla range. Hence the use of electric fields seems not to offer any advantages.

\subsubsection{Resonant polarisation rotation}
\label{sec:ResonantPolarisationRotation}

In 1983 Cruise~\cite{Cruise1} showed that a GW could induce a rotation of the plane of polarisation in electromagnetic waves in certain geometries, some of which might be relevant astronomically. In 2000 the idea of resonant polarization rotation was extended~\cite{Cruise2} to a situation in which the electromagnetic wave was a circulating wave in a microwave waveguide ring. The original effect was amplified by the (potentially significant) quality factor of the waveguide ring. A proof of concept apparatus was constructed by Cruise and Ingley~\cite{Cruise3, Cruise:2006zt}. Such a device would be narrowband with a sensitivity $\sqrt{S_n} \sim 10^{-14}\,\text{Hz}^{-1/2}$ at frequencies of 100\,MHz. It is difficult to see the sensitivity of this scheme for GW detection increasing very far beyond the published value.

\subsubsection{Heterodyne enhancement of magnetic conversion}
\label{sec:AmplificationMethods}

Li and co-workers~\cite{Li:2004df, Li:2006sx, Baker:2008zzb, Li} have suggested enhancing the conversion efficiency of magnetic conversion detectors such as those discussed in Sec.~\ref{sec:inv_gertsenshtein}, {see also Ref.~\cite{Ringwald:2020ist} for a recent discussion}. This proposal has been specifically aimed at the detection of cosmological (relic) signals of a stochastic nature and with dimensionless amplitudes in the range $h_{c,\rm sto} \sim (10^{-30} - 10^{-26})$ at $5 \, \text{GHz}$, about the highest signal consistent with the BBN limit. The conversion from GW to electromagnetic wave is enhanced by seeding the conversion volume with a locally generated electromagnetic wave at the same frequency as that being searched for. In conditions in which the gaussian local oscillator beam is parallel to the incoming signal and at right angles to the static magnetic field, an additional beam of electromagnetic waves is generated by the conversion process, travelling at right angles to the incoming beam and the locally generated beam. The technical challenge is then to distinguish this perpendicular beam of, say, 800 photons/second from the locally generated beam at the same frequency and carrying $10^{24}$ photons/second at frequencies of several GHz~\cite{WOODS201266}. This demands a geometric purity in the Gaussian beam of better than $10^{-21}$, far beyond the current state of the art. The authors have proposed this interesting idea over many years but the lack of laboratory results on the performance of the necessary subsystems leaves the feasibility of this concept an open question.

\subsubsection{Bulk acoustic wave devices}
\label{sec:BAW}

\textit{Bulk acoustic wave devices} are one of the pillars of frequency control and frequency metrology~\cite{ScRep}. In its simplest form, a piece of piezoelectric material is sandwiched between two electrodes, converting the acoustic waves inside the material into electrical signals. With its relatively compact size and robustness, this technology gives one of the best levels of frequency stability near one second of integration time. More recently it was demonstrated that quartz bulk acoustic wave devices exhibit extremely high-quality factors (up to $8 \times 10^9$) at cryogenic temperatures for various overtones of the longitudinal mode covering the frequency range $(5 - 700)\,$MHz~\cite{ScRep,quartzPRL}. For this reason it was proposed to use the technology for various tests of fundamental physics~\cite{ScRep} such as Lorentz invariance tests~\cite{PhysRevX.6.011018}, quantum gravity research~\cite{PhysRevD.100.066020} and search for high-frequency GWs~\cite{Goryachev:2014ac}. For the latter purpose, a bulk acoustic wave device represents a resonant mass detector whose vibration could be read through the piezoelectric effect and a Superconducting Quantum Interference Devices (SQUIDs). The approach has the following advantages: highest quality factor (high-sensitivity), internal (piezoelectric) coupling to SQUIDs~\cite{Goryachev:2014ab}, allows parametric detection methods, large number of sensitive modes ($>100$) in a single device, modes scattered over wide frequency range ($1-700$)\,MHz, well-established and relatively inexpensive technology (mass production), high-precision (insensitive to external influences such as seismic vibration and temperature fluctuations). On the other hand, it is shown that at low temperatures identical devices demonstrate significant dispersion in mode frequencies, thus, showing low accuracy. The level of sensitivity of bulk acoustic wave detectors is estimated at the level of $\sqrt{S_n} \simeq 2\times 10^{-22}\, \text{Hz}^{-1/2}$ subject to the mode geometry~\cite{Goryachev:2014ac}. With additional investment into research and development, this level can be improved and the frequency range extended down to hundreds of kHz range.

A search for high-frequency GWs with a single bulk acoustic wave devices and two modes at 4\,K has been running in the University of Western Australia since November 2018.

\subsubsection{Superconducting rings}
\label{sec:SuperconductingRings}

The quantum properties of vortices in superfluids may interact with spin components of the GWs. In addition, an extension of the electromagnetic impedance at a boundary in the case of GWs in a superconducting fluid suggests that the impedance mismatch well-known in classical bar detector theory is much reduced for a GW arriving at a boundary in a superconductor, essentially creating a very efficient mirror that could be used as a building block for an interferometer or Sagnac ring. Anandan and Chiao~\cite{Anandan, Chiao:2002nv} proposed a new detector format which utilises these putative principles, reaching a sensitivity of $h_{0, n, \rm mono} \sim 10^{-31}$. Resonant operation will restrict the bandwidth in the GHz range. A good review of the issues surrounding the interaction of mesoscopic quantum systems with gravity was prepared on an European Space Agency contract by Kiefer and Weber~\cite{Kiefer}. This review casts doubt on some of the assumptions made by Anandan and Chiao.

\subsubsection{GW deformation of microwave cavities}
\label{sec:HighQCavities}

Caves~\cite{Caves} published a theoretical study of a microwave cavity with a high mechanical quality factor. Mechanical deformation of the cavity by a GW coupled two of the cavity's resonant microwave modes and transferred the electromagnetic excitation to a previously unexcited cavity mode. Reece et al.~\cite{Reece} built a similar system with a higher resonant frequency of 1\,MHz and one operating at $10 \, \text{GHz}$~\cite{Reece:1982sc}, while Pegoraro et al.~\cite{Pegoraro} designed a system with a sharp resonance at about 1\,GHz. These schemes certainly offer some sensitivity in the frequency range above 1\,GHz but that is limited to around $h \sim 10^{-21}$ by the thermal noise in the microwave sensors. Even at cryogenic temperatures the sensitivity will be many orders of magnitude away from the level required for detecting cosmological sources. The bandwidth of this scheme is nominally limited by the bandwidth of the cavity, up to effects like detection or electronics gains and noises.

\subsubsection{Graviton-magnon resonance}
\label{sec:GravitonMagnonResonance}

As pointed out in Ref.~\cite{Ito:2019wcb}, a GW passing through a ferromagnetic insulator can resonantly excite magnons (collective excitations of electron spins), similar to the excitations of phonons in resonant bar detectors. The readout is achieved by placing the magnetic sample inside a microwave cavity, coupling the magnon to a photon mode. This idea builds on the technique of ferromagnetic haloscopes proposed to for axion searches~\cite{Crescini2018,Flower:2018qgb}.
The sensitivity of such detector reaches strain of $\sqrt{S_n} \simeq 7.6 \times 10^{-22} \, \text{Hz}^{-1/2}$ at $14$ GHz and $\sqrt{S_n} \simeq 1.2 \times 10^{-20} \, \text{Hz}^{-1/2}$ at $8.2$ GHz. The sensitivity of this approach can be greatly improved by incorporating single frequency counters that are already available. A few orders of magnitude in sensitivity have been shown for axion detection~\cite{PhysRevD.88.035020}.

\subsection{Summary of detector sensitivities}
\label{sec:SummarySensitivities}

In Tab.~\ref{tab:SummarySensitivity} we summarize the existing and proposed technologies for high-frequency GW detection, reporting the corresponding sensitivities. For all experiments that quote their sensitivity in terms of a power spectral density noise $S_n(f)$, a conversion to dimensionless strain $h_{c, n, \rm sto}$ has been performed using Eq.~\eqref{eq:charnoisesto} with 1 year as the observation time and the specified detector bandwidth. For the other detectors, we specify the dimensionless strain variable on a case by case basis: we use $h_{c, n, \rm sto}$ (see Eq.~\eqref{eq:charnoisesto}) for detectors that look for a stochastic signal, while we use $h_{0, n, \rm mono}$ (see Eq.~\eqref{eq:h0mono}) for detectors that look for a monochromatic GW. In the case of microwave cavities (see Sec.~\ref{sec:HighQCavities}) we denote the sensitivity by $h$ as it can refer either to bursts or to long duration signals, and we refer the reader to the original papers for the details, while we report the best sensitivity estimates. We also specify whether each experiment has already been built, is under construction, is only devised or only the physical mechanism has been identified (theory). {The sensitivity values labeled by an asterisk refer to the planned sensitivities that will be achieved by the proposed future improvements of the currently built setups.} Note that a square bracket in the frequency column refers to the bandwidth of the detector, while a round bracket refers to the range of frequencies\footnote{The range of center frequencies $f_\mathrm{center}$ over which the technical concept can be used, either by building the instrument at the desired operating frequency, or by scanning the frequency as in the case of optically levitated sensors. Not to be confused by the single-shot frequency range spanned by each detector.} that can be covered by the detector itself. We report the bandwidths used in Tab.~\ref{tab:SummarySensitivity} to obtain dimensionless strain from PSD: Mini-GRAIL had a bandwidth $\Delta f  \simeq 30 \, \text{Hz}$~\cite{Gottardi:2007zn}; the Schenberg antenna had a bandwidth $\Delta f \simeq 50 \, \text{Hz}$~\cite{Aguiar:2010kn}; the $0.75$ m interferometer has bandwidth $\Delta f \simeq 2 \, \text{kHz}$~\cite{akutsu}\footnote{Note that, despite the $0.75$ m interferometer~\cite{akutsu} is sensitive to frequencies around $100 \, \text{MHz}$, we do not see any reason that prevents this concept to be extended in the range kHz-$100\,$MHz. This is the reason why we covered this frequency range in Fig.~\ref{fig:hc_coherent_summary} and Fig.~\ref{fig:hc_stochastic_summary}.}; the optically levitated sensors have a bandwidth \(\Delta f = f/10\)
~\cite{arvanitaki:2016gw,aggarwal2020searching}; $\Delta f \simeq (10-50) \, \text{kHz}$ for Cruise's and Ingley's detector~\cite{Cruise3, Cruise:2006zt}; for enhanced magnetic conversion the bandwidth is $\Delta f \simeq 1 \, \text{Hz}$~\cite{Li}; bulk acoustic wave resonators have bandwidth $\Delta f = f/10^8$, where $f$ is in the ranges reported in Tab.~\ref{tab:SummarySensitivity}
~\cite{Goryachev:2014ab}; the bandwidth for Pegoraro's detector is $\Delta f \simeq 1 \, \text{Hz}$. For references that specify their detector sensitivity in dimensionless strain without specifying the exact form of the dimensionless strain, the sensitivities are labeled as just \(h\).

\begin{table}[!htbp]
\begin{footnotesize}
\centerline{
\begin{tabular}{|@{}c@{}||@{}c@{}|@{}c@{}|@{}c@{}|}
\hline
\begin{tabular}{c} \centered{Technical concept} \end{tabular} & \begin{tabular}{c}  Operational\\ Frequency \end{tabular} &
\begin{tabular}{c} Proposed sensitivity \\ (dimensionless) \end{tabular} & \begin{tabular}{c} Proposed sensitivity \\ $\sqrt{S_n(f)}$ \end{tabular} \\[1.4ex]
\hline
\hline
\begin{tabular}{c} \textbf{Spherical resonant mass}, Sec.~\ref{sec:spheres}~\cite{Forward1971} \end{tabular} &  &  & \\[1.2ex]
\cline{1-4}
\begin{tabular}{c} \centered{Mini-GRAIL (built)~\cite{Gottardi:2007zn}} \end{tabular} & \begin{tabular}{c} $2942.9$ Hz \end{tabular} & \begin{tabular}{c} \centered{$10^{-20}$} \\ \centered{$2.3 \cdot 10^{-23}\,(*)$} \end{tabular} & \begin{tabular}{c} $5 \cdot 10^{-20} \, \text{Hz}^{-\frac{1}{2}}$ \\ \centered{$10^{-22} \, \text{Hz}^{-\frac{1}{2}} \,(*)$} \end{tabular} \\[1.4ex]
\cline{1-4}
\begin{tabular}{c} \centered{Schenberg antenna (built)~\cite{Aguiar:2010kn}} \end{tabular} & \begin{tabular}{c} $3.2$ kHz \end{tabular} & \begin{tabular}{c} $2.6\cdot 10^{-20}$ \\ $2.4\cdot 10^{-23}\,(*)$ \end{tabular} & \begin{tabular}{c} $1.1 \cdot 10^{-19} \, \text{Hz}^{-\frac{1}{2}}$ \\ $10^{-22} \, \text{Hz}^{-\frac{1}{2}}\,(*)$ \end{tabular} \\[1.4ex]
\hline
\hline
\begin{tabular}{c} \centered{\textbf{Laser interferometers}} \end{tabular} &   &   & \\[1.4ex]
\cline{1-4}
\begin{tabular}{c} \centered{NEMO (devised), Sec.~\ref{sec:ozgrav}~\cite{ozhf, Bailes:2019oma}} \end{tabular} & \begin{tabular}{c} $[1-2.5]$ kHz \end{tabular} & \begin{tabular}{c} $9.4 \cdot 10^{-26}$ \end{tabular} & \begin{tabular}{c} $10^{-24} \, \text{Hz}^{-\frac{1}{2}}$ \end{tabular} \\[1.4ex]
\cline{1-4}
\begin{tabular}{c} \centered{$0.75$ m interferometer (built), Sec.~\ref{sec:100MHzInterferometers}~\cite{akutsu, Nishizawa:2007tn}} \end{tabular} & \begin{tabular}{c} $100$ MHz \end{tabular} & \begin{tabular}{c} $7 \cdot 10^{-14}$ \\ $2 \cdot 10^{-19}\,(*)$ \end{tabular} & \begin{tabular}{c} $10^{-16}~\text{Hz}^{-\frac{1}{2}}$ \\ $10^{-20} \, \text{Hz}^{-\frac{1}{2}} \, (*)$ \end{tabular} \\[1.4ex]
\cline{1-4}
\begin{tabular}{c} \centered{Holometer (built), Sec.~\ref{sec:100MHzInterferometers}~\cite{PhysRevD.95.063002}} \end{tabular} & \begin{tabular}{c} $[1-13]$ MHz \end{tabular} & \begin{tabular}{c}  $2.5 \cdot 10^{-18} - 2.4 \cdot 10^{-19}$ \end{tabular} & \begin{tabular}{c}  $10^{-21} \, \text{Hz}^{-\frac{1}{2}}$ \end{tabular} \\[1.4ex]
\hline
\hline
\begin{tabular}{c} \centered{\textbf{Optically levitated sensors}, Sec.~\ref{sec:OpticallyLevitatedSensors}~\cite{aggarwal2020searching}} \end{tabular} &  &  &  \\[1.4ex]
\cline{1-4}
\begin{tabular}{c} \centered{$1$-meter prototype (under construction)} \end{tabular} & \begin{tabular}{c} $(10-100)$ kHz \end{tabular} & \begin{tabular}{c} $2.4\cdot 10^{-20} - 4.2 \cdot 10^{-22}$ \end{tabular} & \begin{tabular}{c}  $(10^{-19}-10^{-21}) \, \text{Hz}^{-\frac{1}{2}}$ \end{tabular} \\[1.4ex]
\cline{1-4}
\begin{tabular}{c} \centered{$100$-meter instrument (devised)} \end{tabular} & \begin{tabular}{c} $(10-100)$ kHz \end{tabular} & \begin{tabular}{c} $2.4\cdot 10^{-22} - 4.2 \cdot 10^{-24}$ \end{tabular} & \begin{tabular}{c} $(10^{-21}-10^{-23}) \, \text{Hz}^{-\frac{1}{2}}$ \end{tabular} \\[1.4ex]
\cline{1-4}
\hline
\hline
\begin{tabular}{c} \centered{\textbf{Inverse Gertsenshtein effect}, Sec.~\ref{sec:inv_gertsenshtein}} \end{tabular} &  &  &  \\[1.4ex]
\cline{1-4}
\begin{tabular}{c} \centered{GW-OSQAR II (built)~\cite{Ejlli2019}} \end{tabular} & \begin{tabular}{c} $(2.7-14) \cdot 10^{14}$ Hz \end{tabular} & \begin{tabular}{c}  $h_{c, n, \rm sto} \simeq 8 \cdot 10^{-26}$ \end{tabular} & \begin{tabular}{c} $\times$ \end{tabular} \\[1.4ex]
\cline{1-4}
\begin{tabular}{c} \centered{GW-CAST (built)~\cite{Ejlli2019}} \end{tabular} & \begin{tabular}{c} $(5-12) \cdot \, 10^{18}$ Hz \end{tabular} & \begin{tabular}{c}  $h_{c, n, \rm sto} \simeq 7 \cdot 10^{-28}$ \end{tabular} & \begin{tabular}{c} $\times$ \end{tabular} \\[1.4ex]
\cline{1-4}
\begin{tabular}{c} \centered{GW-ALPs II
(devised)~\cite{Ejlli2019}} \end{tabular} & \begin{tabular}{c} $\sim 10^{15}$ Hz \end{tabular} & \begin{tabular}{c} $h_{c, n, \rm sto} \simeq 2.8 \cdot 10^{-30}$ \end{tabular} & \begin{tabular}{c} $\times$ \end{tabular} \\[1.4ex]
\hline
\hline
\begin{tabular}{c} \centered{\textbf{Resonant polarization rotation}, Sec.~\ref{sec:ResonantPolarisationRotation}~\cite{Cruise1}} \end{tabular} &  &  &  \\[1.4ex]
\cline{1-4}
\begin{tabular}{c} \centered{Cruise's detector (devised)~\cite{Cruise2}} \end{tabular} & \begin{tabular}{c} $(0.1 - 10^5)\,\text{GHz}$ \end{tabular} & \begin{tabular}{c} $h_{0, n, \rm mono} \simeq 10^{-18}$ \end{tabular} & \begin{tabular}{c} $\times$ \end{tabular} \\[1.4ex]
\cline{1-4}
\begin{tabular}{c} \centered{Cruise \& Ingley's detector (prototype)~\cite{Cruise3, Cruise:2006zt}} \end{tabular} & \begin{tabular}{c} $100$ MHz \end{tabular} & \begin{tabular}{c} $8.9 \cdot 10^{-14}$ \end{tabular} & \begin{tabular}{c} $10^{-14} \, \text{Hz}^{-\frac{1}{2}}$ \end{tabular} \\[1.4ex]
\hline
\hline
\begin{tabular}{c} \centered{\textbf{Enhanced magnetic conversion}} \\ \centered{(theory), Sec.~\ref{sec:AmplificationMethods}~\cite{Li}} \end{tabular} & \begin{tabular}{c} $\sim 10$ GHz \end{tabular} & \begin{tabular}{c} $h_{c, n, \rm sto} \simeq 10^{-30} - 10^{-26}$ \end{tabular} & \begin{tabular}{c} $\times$ \end{tabular} \\[1.4ex]
\hline
\hline
\begin{tabular}{c} \centered{\textbf{Bulk acoustic wave resonators}} \\ \centered{(built), Sec.~\ref{sec:BAW}~\cite{Goryachev:2014ab, Goryachev:2014ac}} \end{tabular} & \begin{tabular}{c} $(\text{MHz} - \text{GHz})$ \end{tabular} & \begin{tabular}{c}  $4.2\cdot10^{-21} - 2.4 \cdot 10^{-20}$ \end{tabular} & \begin{tabular}{c}  $10^{-22} \, \text{Hz}^{-\frac{1}{2}}$ \end{tabular} \\ [1.4ex]
\hline
\hline
\begin{tabular}{c} \centered{\textbf{Superconducting rings}, (theory), Sec.~\ref{sec:SuperconductingRings}~\cite{Anandan, Chiao:2002nv}} \end{tabular} & \begin{tabular}{c} $10$ GHz \end{tabular} & \begin{tabular}{c} $h_{0,n, \rm mono} \simeq 10^{-31}$ \end{tabular} & \begin{tabular}{c} $\times$ \end{tabular} \\[1.4ex]
\hline
\hline
\begin{tabular}{c} \centered{\textbf{Microwave cavities}, Sec.~\ref{sec:HighQCavities}} \end{tabular} &  &  &  \\[1.4ex]
\cline{1-4}
\begin{tabular}{c} \centered{Caves' detector (devised)~\cite{Caves}} \end{tabular} & \begin{tabular}{c} $500$ Hz \end{tabular} & \begin{tabular}{c} $h \simeq 2 \cdot 10^{-21}$ \end{tabular} & \begin{tabular}{c} $\times$ \end{tabular} \\[1.4ex]
\cline{1-4}
\begin{tabular}{c} \centered{Reece's 1st detector (built)~\cite{Reece}} \end{tabular} & \begin{tabular}{c} $1$ MHz \end{tabular} & \begin{tabular}{c} $h \simeq 4 \cdot 10^{-17}$ \end{tabular} & \begin{tabular}{c} $\times$ \end{tabular} \\[1.4ex]
\cline{1-4}
\begin{tabular}{c} \centered{Reece's 2nd detector (built)~\cite{Reece:1982sc}} \end{tabular} & \begin{tabular}{c} $10$ GHz \end{tabular} & \begin{tabular}{c} $h \simeq 6 \cdot 10^{-14}$ \end{tabular} & \begin{tabular}{c} $\times$ \end{tabular} \\[1.4ex]
\cline{1-4}
\begin{tabular}{c} \centered{Pegoraro's detector (devised)~\cite{Pegoraro}} \end{tabular} & \begin{tabular}{c} $(1-10)$ GHz \end{tabular} & \begin{tabular}{c} $h \simeq 10^{-23}$ \end{tabular} & \begin{tabular}{c} $\times$ \end{tabular} \\[1.4ex]
\hline
\hline
\begin{tabular}{c} \centered{\textbf{Graviton-magnon resonance}} \\ \centered{(theory), Sec.~\ref{sec:GravitonMagnonResonance}~\cite{Ito:2019wcb}} \end{tabular} & \begin{tabular}{c} $(8-14)$ GHz \end{tabular} & \begin{tabular}{c} $1.1 \cdot 10^{-12} - 1.3 \cdot 10^{-13}$ \end{tabular} & \begin{tabular}{c} $(10^{-22} - 10^{-20}) \, \text{Hz}^{-\frac{1}{2}}$ \end{tabular} \\[1.4ex]
\cline{1-4}
\end{tabular}}
\end{footnotesize}
\caption{\label{tab:SummarySensitivity} Summary of existing and proposed detectors with their respective sensitivities. See Sec.~\ref{sec:SummarySensitivities} for details.}
\end{table}


\subsection{Cross correlation detectors}
\label{sec:crosscorrelation}

For a coalescing binary the information about the waveform of GWs is available. The best approach to the detection problem is to use this information by projecting the observed data over the set of expected signals. Usually this set can be parameterized by a small number of parameters, varying in some allowed range. A `scalar product' between data and the set of `templates' is evaluated, using its maximum as a detection statistics.
It should be noted that the set of expected signals does not have a linear space structure, in the sense that the linear combination of two possible signals is not generally speaking a possible signal. For this reason the computational cost of a search over a `template bank' grows very fast with the number of free parameters.

When the number of parameters is large, or when a parameterization of the waveform of the expected signal is not possible at all, other detection methods must be used. In the present context this is the case for several cosmological processes, which are expected to produce a GW signal that can be described as an overlap of a very large number of contributions. There is not a waveform here, the expected signal is a stochastic process and the best approach for the detection is the cross correlation one.

\subsubsection{Relic gravitational radiation}

The detection of the CMB in 1965 by Penzias and Wilson gave the first experimental insight into the properties of `relic' radiation, the remnants currently observable of the Big Bang. The CMB is a stationary, stochastic radiation field, basically isotropic down to levels as low as $10^{-5}$ with a Gaussian distribution.

It is natural to suppose that a useful starting point in planning GW observations of a relic radiation would be to assume that cosmologically-sourced GWs can be modeled by a stochastic field $h_{ij}(\boldsymbol{x},t)$ with relatively simple properties.
In the spacetime volume of a given experiment, we can describe our background as a superposition of plane waves
\begin{equation}
h_{ij}(\boldsymbol{x},t) = \sum_P \int \frac{d^3 k}{(2\pi)^3} \tilde{h}_P(\boldsymbol{k}) \varepsilon_{ij}^P(\hat{k})
e^{i\boldsymbol{k}\boldsymbol{x}-i\omega_k t}
+\text{c.c} \,.
\end{equation}
Here $P$ labels the polarization degrees of freedom of the field, whose number can depend on the considered theory of gravitation. The `parameters' of the signal are the amplitudes $h_P(\boldsymbol{k})$ introduced in Sec.~\ref{sec:th}, one for each mode of the field. When these must be considered stochastic variables, the relic gravitational radiation field is completely described by their (joint) probability distribution, and is called GW stochastic background.

For a Gaussian stochastic background this joint probability distribution is Gaussian, and is completely determined by the second order expectation value
\begin{equation}
	\left< \tilde{h}_A(\boldsymbol{p}) \tilde{h}_B(\boldsymbol{q})  \right> = \Gamma_{AB}\left(\boldsymbol{p},\boldsymbol{q}\right) \,.
\end{equation}
Further assumptions lead to the simplification of this general expression. For example, stationarity in a given reference frame requires the expectation value $\langle h_{ij}(\boldsymbol{x},t)h_{ij}(\boldsymbol{y},t^\prime)\rangle$ to be a function of $t-t^\prime$ only, and as a consequence the correlation between $h_A(\boldsymbol{p})$ and $h_B(\boldsymbol{q})$ can be non-zero only when $\omega_p=\omega_q$. Stationarity in every reference frame implies homogeneity, and only correlations with $\boldsymbol{p}=\boldsymbol{q}$ are allowed.

Each stochastic model can have its peculiar signature (see also Sec.~\ref{sec:th}): in the simplest stationary, isotropic and Gaussian model a parameterization of the model can be given in term of an array of functions which are connected to contributions
\begin{equation}
	 \Gamma_{AB}\left(\boldsymbol{p},\boldsymbol{q}\right) \propto \frac{1}{f^3} \delta^2\left(\hat{\boldsymbol{p}},\hat{\boldsymbol{q}}\right) \delta(\omega_p-\omega_q) \Omega_{\rm GW}^{AB}\left(\omega_p\right) \,,
 \end{equation}
which is a generalization of Eq.~(\ref{eq:Sh}) and allows for a non trivial polarization structure.

If the radiation is stationary, a temporal signature which could be exploited by a single-instrument detection procedure is not available. Moreover, isotropy and homogeneity do not allow for a signal modulation which could be obtained in principle by changing the orientation or the position of the detector.\footnote{Note however that the motion of the detector with respect to the cosmic rest frame of a cosmological stochastic background breaks isotropy. This yields a (weak) signal modulation and can be exploited to extract polarization information from the stochastic background of GWs~\cite{Seto:2006hf,Seto:2006dz,Domcke:2019zls}.}

It could still be possible to detect the stochastic background as an `excess noise' in the apparatus. However in order to do that the amplitude of the signal must be large enough to make it evident given a theoretical estimate of the noise budget, which is always uncertain. This means that the strategy for detection will necessarily be different from the strategy for discrete source detection.

The most obvious approach is the use of spatial correlations. If a detector is to be developed for the detection of GW relic radiation then a decision to operate it as a correlation detector will have a far reaching influence on many aspects of its design.

An excellent review of GW relic radiation and appropriate methods of detection, in the context of HF band, has been published by Allen \cite{Allen} and important properties of correlation detectors have been explored by Michelson~\cite{Michelson}. The basic principle is to compare the signal from two detectors. This is comparing a random signal with another stationary, stochastic, isotropic, gaussian signal from the same source. Similar to template matching as a means of detecting discrete sources, in this case the template itself is random and that affects the statistical gain from performing a cross correlation between two detectors.

\subsubsection{Properties of correlation detectors}

At the simplest level the outputs $s_i(t)$ of two detectors can be written as
\begin{align}
	s_1(t) &= h_1(t) + n_1(t) \,,\nonumber \\
	s_2(t) &= h_2(t)+  n_2(t) \,,
\end{align}
where $h_{1,2}$ are the stochastic signals and $n_{1,2}$ the noises. We suppose both the signals and the noises to have typical amplitudes $h_i$, $n_i$ and typical bandwidth $\Delta f$.
If we evaluate the simple correlation over a time $T\gg \Delta f^{-1}$ between the two outputs
\begin{align}
	\hat{Y} = \int_t^{t+T} s_1(t') s_2(t') dt' \,,
\end{align}
we find an estimator which is distributed as a Gaussian variable for large enough $T$. Its mean is given by
\begin{align}
	\left<\hat{Y}\right> = \int_t^{t+T} \left<h_1(t') h_2(t')\right> dt' \propto C^{h_1 h_2}(0) T \,,
\end{align}
where the temporal cross-correlation $C^{h_1 h_2}$ is defined as
\begin{equation}
C^{xy}(\tau) = \langle x(t) x(t+\tau) \rangle \,,
\end{equation}
and we supposed that the two noises are uncorrelated. The variance of $\hat Y$ is given by
\begin{align}
	\left<\hat{Y}^2\right>-\left<\hat{Y}\right>^2 & = \int_t^{t+T} dt' \int_t^{t+T} dt'' C^{n_1 n_1}(t'' - t') C^{n_2 n_2}(t''-t') \simeq \nonumber \\
	&\simeq \int_{-\infty}^{\infty} d\tau C^{n_1 n_1}(\tau) C^{n_2 n_2}(\tau) \simeq \nonumber \\
	&\simeq T C^{n_1 n_1}(0) C^{n_2 n_2}(0) \frac{1}{\Delta f} \,,
\end{align}
and we see that the signal-to-noise ratio
\begin{align}
	\frac{\left<\hat{Y}\right>}{\sqrt{\left<\hat{Y}^2\right>-\left<\hat{Y}\right>^2}} \propto \frac{C^{h_1 h_2}(0)}{\sqrt{C^{n_1 n_1}(0) C^{n_2 n_2}(0)}} \sqrt{T\Delta f} \,,
\end{align}
increases as the square root of the measurement time. The minimum detectable energy density $\Omega_{\rm GW}$ is
\begin{equation}
\Omega^{\rm min}_{\rm GW} \propto \sqrt{\frac{C^{n_1 n_1}(0) C^{n_2 n_2}(0)}{T \Delta f}} \,,
\end{equation}
 and decreases with the square root of $T$. Of course the minimum detectable signal amplitude is proportional to $\sqrt{\Omega^{\rm min}_{\rm GW}}$, so it decreases much slowly.

\subsubsection{The overlap function}

In the previous, very simple analysis we neglected all the effects that can be adsorbed in a proportionality factor.  For example when we evaluated the signal, we wrote
\begin{equation}
\left<h_1(t) h_2(t)\right> = K h_1 h_2 \,,
\end{equation}
implicitly assuming that $K=O(1)$. As a matter of fact the correlation between the signals coupled to two different detectors can be reduced by several effects.
The quantity of interest can always been obtained by the correlation between the gravitational strain at two different points
\begin{equation}
\left<\tilde{h}_{ij}^*(\boldsymbol{x},\omega) \tilde{h}_{k\ell}(\boldsymbol{y},\omega^\prime)\right> \propto
\sum_{P,P^\prime} \int d\hat{\Omega}_k d\hat{\Omega}_{k^\prime} \left<\tilde{h}_P^*(\hat{k},\omega)\tilde{h}_{P^\prime}(\hat{k}^\prime,\omega^\prime) \right> \varepsilon_{ij}^P(\hat{k}) \varepsilon_{k\ell}^{P^\prime}(\hat{k}^\prime)
e^{-i\omega(\hat{\boldsymbol{k}}\boldsymbol{x}-\hat{\boldsymbol{k}^\prime}\boldsymbol{y})}
\end{equation}
after a contraction on appropriate tensors which describe the detectors. Without entering in the details of the calculation, we see that the value of the previous correlation is influenced by two effects.

First of all, the detectors will not be in the same position. In this case the phase factor in the integral will oscillate, and the correlation will be reduced. It is clear that this reduction will start when the separation between the two detectors will be larger than the wavelength of the field at the frequency of interest.  In the very high-frequency regime, this will happen at very small separations, $d \gtrapprox 1/(2\pi f)$.

A further reduction of the correlation can be generated by two detectors which are coupled differently to the modes, for example because they are oriented differently.

The reduction of correlation is quantified by the so--called `overlap function' $\gamma (f)$, a frequency dependent factor with modulus always less than 1, which is simply the coherence between the two signal of interest. In a quantitative analysis $\gamma(f)$ must be included so as to diminish the signal correlation and increase the minimum detectable amplitude by a factor $\frac{1}{\gamma(f)}$. Michelson~\cite{Michelson} has worked through the derivation of $\gamma(f)$ and Allen \cite{Allen} has outlined the process of optimising the detection efficiency by optimal filtering in the time domain for two detectors with arbitrary separation and orientation.

\subsubsection{Exploiting $\gamma(f)$ at very high-frequencies}

One possibility that arises at high-frequencies due to the small wavelength of the radiation being studied is to construct laboratory scale detectors which in principle could be moved relative to one another, hence changing the value of the overlap function $\gamma$. In this way the correlation of the signal can be modulated, and a detection of this modulation pattern could provide credible evidence of a relic radiation detection. It is interesting to note that the overlap reduction function depends on the polarization structure of the field. This means that by looking at the modulation pattern it is in principle possible to disentangle non standard polarizations coming from extended gravitational theories.

\subsubsection{Signal switching}
\label{subsec:switch}

The prospect of a new detection technology for high-frequency GW detection opens a window for a slightly novel form of correlation detector. Instead of correlating the output signal with that of a similar detector, it may be possible to turn off the sensitivity of a single detector to GWs without affecting its other performance properties. If the temporal pattern of switching on and off can be seen in the signal output at a statistically significant level then a credible claim for a detection might be made.

An opportunity of this kind presents itself for correlation detection in the case of magnetic conversion detectors (see below), in which case signal switching could also be achieved by modulating the amplitude of the field and its direction. The devices are arranged to enable the conversion of the GW to an electromagnetic one within an electromagnetic cavity crossed by a strong magnetic field. The electromagnetic signal produced is proportional to the length of the cavity in GW wavelengths and the cavity dimensions transverse to its length are critical in determining the efficiency of the process. This is because the generated electromagnetic wave finds itself travelling in an electromagnetic waveguide, with a phase velocity greater than that of the GW. To make the conversion efficient these two phase velocities must be sufficiently close that, in the whole length of the detector, only a very small phase difference develops between the two waves. If this is not the case then the conversion process is essentially rendered inoperative. Current thoughts on slowing the electromagnetic wave in the cavity include the introduction of a gas into the cavity and adjusting the refractive index at the electromagnetic frequency by means of the pressure to equalise the phase velocities. Because the conversion is so inefficient if this is not done, controlling the gas pressure can essentially switch the sensitivity to GWs off and on with very minimal disturbance to its other operating parameters and hence to the output signal. The detector output can then be correlated with the detector operating pattern. Statistically, this is a more effective correlation process than described above between two similar detectors because in this case the signal is compared with an a priori determined template and not a random template. The minimum detectable signal in this case is $\propto T^{-\frac{1}{2}}$ allowing a faster gain in sensitivity with time.

\subsubsection{Issues related to data acquisition and long term storage}

To detect correlated periodic events in nearby detectors at frequencies around 1 MHz at an signal-to-noise ratio of 8, systematic errors related to timing should be of the order of 20 ns. This necessitates the need for a careful understanding of the various factors that contribute to errors arising from timing calibration. From the hardware side, low noise amplifiers, anti-aliasing electronics, etc. add delay to the data acquisition system. Quantization errors arising from the analog-to-digital converters add further delay and to minimize the effect, their sampling period would have to be made lower than the desired timing resolution. At such sampling rates, making use of super-conducting oversampling ADCs which achieve high dynamic ranges over narrow frequency ranges by pushing quantization noise outside the band-of-interest could turn out to be viable option.

Both stochastic and continuous wave analysis rely on cross-correlation based search strategies and make use of year long data-sets for their analysis. Moving from current audio-band frequencies to megahertz regime can easily scale up the storage requirements to few petabytes of data. However it must be noted that the increase of storage is not proportional to the frequency of interest, but on the bandwidth, as the typical observation frequency can always scaled down with an appropriate heterodyne technique.

This calls for real-time analysis as proposed in SKA where the raw data is discarded after the low latency retrieval of relevant information. Advantages of using cloud and web 2.0 technologies for the collaborative development of the various data analysis pipelines used for signal detection and parameter estimation also needs to be investigated. Making use of data folding techniques based on inherent symmetries, such as Earth's siderial day rotation has been shown to decrease data volume in stochastic background searches at audio-band frequencies~\cite{PhysRevD.92.022003}. Stacking years long data into a single day while preserving all the statistical properties would enable us to carry out the final analysis on personal computing devices.


\section{Discussion and conclusions}
\label{sec:conclusion}

The search for high-frequency gravitational waves is a promising and challenging search for new physics. It provides a unique opportunity to test many theories beyond the Standard Model that could not be tested otherwise.

Various models proposed to address open questions in particle physics and cosmology predict a gravitational wave signal in the frequency range $f \simeq (10^3 -  10^{10})\,$Hz, which could be a coherent signal - e.g.\ from mergers of compact objects or from axion superradiance around black holes - or stochastic - e.g.~originating from certain models of cosmic inflation, from a phase transition in the very early Universe or from oscillons, evaporating primordial black holes, etc. See Figs.~\ref{fig:hc_stochastic_summary} and~\ref{fig:hc_coherent_summary} as well as Tab.~\ref{tab:summary-coherent} and Tab.~\ref{tab:summary-stochastic} for an overview. Many of these models can lead to relatively large signals, corresponding to an ${\cal O}(1)$ fraction of energy density in the early Universe converted to gravitational waves. This energy is red-shifted in the expanding Universe, rendering even these strong signals challenging to detect today. Moreover, in many cases the amplitude of the signal depends sensitively on the model parameters and may be significantly lower in large parts of the model parameter space.

The high-frequency band comes with particular challenges and opportunities. High-frequency gravitational waves carry a high-energy density, implying that bounds provided by cosmology on the fraction of energy contained in gravitational waves translate to stringent bounds on the characteristic gravitational wave strain. This poses a severe challenge for detection, since the magnitude of any observable effects is typically governed by the strain and not by the energy density. This renders the detection of cosmological sources of high-frequency gravitational waves much more challenging than comparable searches at lower frequencies. On the other hand, the lack of known astrophysical gravitational wave sources in this frequency range poses a unique opportunity for foreground free searches of new physics.

At the moment, there is no general consensus on the most promising detection strategy in this frequency band, though many proposals have been put forward in the past decades. The proposals that we are aware of are summarized in Tab.~\ref{tab:SummarySensitivity}, together with their frequency and sensitivity range. We emphasize that the same sensitivity (in terms of characteristic strain) at a higher frequency typically implies a reduced sensitivity to the viable parameter space of a given cosmological source, as discussed above. In this sense, detectors based on magnetic conversion or on the deformation of microwave cavities seem to be particularly promising avenues, though a more careful study of noise levels and of the margin on improvement with foreseeable technology development is needed in many cases.

We hope that this document will stimulate the necessary discussion and we strongly encourage feedback regarding further proposals or critical assessments which we may have missed. We have the ambition to consider this document as a first step towards a coherent international collaboration to seriously consider the search for high-frequency gravitational waves.

None of the proposals listed in this report currently reach the sensitivity needed to probe the new physics outlined above. At best, achievable proposals are still at least six orders of magnitude beyond the required sensitivity. However, we recall that, one hundred years ago, the technological gap in both the LIGO and LISA frequency ranges was of about $16$ orders of magnitude~\cite{Chen:2016isk}. Also, less than 50 years ago, Misner, Thorne and Wheeler, declared that `\textit{such detectors have so low sensitivity that they are of little experimental interest}'~\cite{Misner:1974qy}, referring to laser interferometers. The first laser interferometer gravitational wave detector, built by Robert Forward at Hughes Research Laboratories in the 1970's~\cite{PhysRevD.17.379} had a sensitivity which was eight orders of magnitudes below the design sensitivity of the currently operating LIGO/Virgo/KAGRA detectors. There are currently clear development paths leading to detectors operating at sensitivities of $h_{c, n} \simeq 10^{-26}$ using e.g.\ magnetic conversion (see Sec.~\ref{sec:inv_gertsenshtein}) and several new ideas such as magnon devices and superconducting systems which have received little detailed design development so far.

We therefore take the past history of laser interferometry as an encouraging lesson for the development of gravitational wave detectors in the high-frequency band. The challenges are strong but the opportunities are unique. As a rule of thumb, probing very early epochs of our cosmic history and consequently particle physics at very high-energy scales requires searching for gravitational waves at high-frequencies with correspondingly small experimental devices. As can be seen from Fig.s~\ref{fig:hc_stochastic_summary} and \ref{fig:hc_coherent_summary}, the Ultra High-Frequency band, ranging from the MHz to the GHz is an exciting window to explore fundamental physics up to the grand unification or string theory scales of order $(10^{16}-10^{17})$~GeV, thereby probing our cosmological history right up to the Big Bang. A detection of a gravitational wave signal in this frequency range would be smoking gun signal for new physics, since no known astrophysical processes can generate sizable gravitational wave signals at these frequencies. It would be remarkable if the experimental test of fundamental physics at the highest energies and earliest times in the history of the Universe could eventually be achieved not with huge particle accelerators nor satellite interferometry, but with small size table-top experiments.

This white paper set up the stage for the launch of the Ultra-High-Frequency Gravitational Wave (UHF-GW) initiative\footnote{Check out the \href{http://www.ctc.cam.ac.uk/activities/UHF-GW.php}{\underline{website}} of the initiative.}, that supports the creation of a network of researchers for the development of gravitational wave science in the high-frequency range. One of the goals of the initiative is to stimulate the technological development that is needed to build successful gravitational wave detectors at high-frequency.


\paragraph{Acknowledgements}
We thank all the ICTP staff for providing the excellent conditions that allowed us to organise this workshop, especially to Nadia van Buuren for her efficient and friendly administrative support.
We moreover thank Ken-Ichi Herada, Sotatsu Otabe, Seyed Mohammad Sadegh Movahed, and Masha Baryakhtar for valuable input. We also thank the two referees of Living Reviews in Relativity for their supportive and well-thought reports.\\
The Australian High-Frequency Gravitational Wave Effort is supported by the Australian Research Council Centre of Excellence for Gravitational Wave Discovery (OzGrav), Grant number CE170100004. 
N.A. is supported by NSF grant PHY-1806671 and a CIERA Postdoctoral Fellowship from the Center for Interdisciplinary Exploration and Research in Astrophysics at Northwestern University.
A.B. acknowledges support by the European Research Council (ERC) under the European Union's Horizon 2020 research and innovation programme under grant agreement No. 759253 and by Deutsche Forschungsgemeinschaft (DFG, German Research Foundation) - Project-ID 279384907 - SFB 1245 and - Project-ID 138713538 - SFB 881 (`The Milky Way System', subproject A10).
O.D.A. thanks FAPESP/Brazil (grant numbers 1998/13468-9 and 2006/56041-3) and CNPq/Brazil (grants numbers 306467{\textunderscore}2003{\textunderscore}8, 303310{\textunderscore}2009-0, 307176{\textunderscore}2013-4, and 302841/2017-2).
This project has received funding from the Deutsche Forschungsgemeinschaft under Germany's Excellence Strategy\,--\,EXC 2121 `Quantum Universe'\,--\,390833306 (V.\,D., F.\,M.).
D.G.F. (ORCID 0000-0002-4005-8915) is supported by a Ram\'on y Cajal contract by Spanish Ministry MINECO, with Ref. RYC-2017-23493, and by the grant `SOM: Sabor y Origen de la Materia', from Spanish Ministry of Science and Innovation, under no. FPA2017-85985-P.
A.G. is supported in part by NSF grants PHY-1806686 and PHY-1806671, the Heising-Simons Foundation, the W.M. Keck Foundation, the John Templeton Foundation, and ONR Grant N00014-18-1-2370.
M.G. and M.E.T. were funded by the ARC Centre for Excellence for Engineered Quantum Systems, CE170100009, and the ARC Centre for Excellence for Dark Matter Particle Physics, CE200100008, as well as ARC grant DP190100071. F.M. is funded by a UKRI/EPSRC Stephen Hawking fellowship, grant reference EP/T017279/1. This work has been partially supported by STFC consolidated grant ST/P000681/1. A.R. acknowledges funding from Italian Ministry of Education, University and Research (MIUR) through the `Dipartimenti di eccellenza' project Science of the Universe. S.S.  was supported by MIUR in Italy under Contract(No. PRIN 2015P5SBHT) and ERC Ideas Advanced Grant (No. 267985) `DaMeSyFla'. F.T. acknowledges support from the Swiss National Science Foundation (project number 200020/175502). C.U. is supported by European Structural and Investment Funds and the Czech Ministry of Education, Youth and Sports (Project CoGraDS - CZ.$02.1.01/0.0/0.0/15\_003/0000437$) and partially supported by ICTP.

\appendix

\section{Summary tables}
\label{sec:SummaryTable}

\begin{table}[!ht]
\begin{footnotesize}
\centerline{
\begin{tabular}{|@{}c@{}||@{}c@{}|@{}c@{}|@{}c@{}|}
\hline
\begin{tabular}{c} Source \end{tabular} & \begin{tabular}{c} Typical frequency \end{tabular} & \begin{tabular}{c} Characteristic strain $h_c$ \\ (dimensionless) \end{tabular} \\[1.4ex]
\cline{1-3}
\begin{tabular}{c} Neutron star mergers: \\ binaries \end{tabular} & \begin{tabular}{c} $(1-5) \, \text{kHz}$\end{tabular} & \begin{tabular}{c} $\lesssim 10^{-21}$ \end{tabular} \\[1.4ex]
\cline{1-3}
\begin{tabular}{c} Primordial BH mergers: \\ binaries \end{tabular} & \begin{tabular}{c} $\frac{4400} {(m_1 + m_2)} {\rm Hz}$ \end{tabular} & \begin{tabular}{c} $\lesssim 4.2 \times 10^{-20} \, \left(\frac{\text{Hz}}{f}\right)^{0.7}$ \end{tabular} \\[1.4ex]
\cline{1-3}
\begin{tabular}{c} Primordial BH mergers: \\ capture in haloes \end{tabular} & \begin{tabular}{c} $\frac{4400} {(m_1 + m_2)} {\rm Hz}$ \end{tabular} & \begin{tabular}{c} $\lesssim 6.1 \times 10^{-20} \, \left(\frac{\text{Hz}}{f}\right)$ \end{tabular} \\[1.4ex]
\cline{1-3}
\begin{tabular}{c} Exotic compact objects \end{tabular} & \begin{tabular}{c} $C^{3/2} \left(\frac{6 \times 10^{-3} M_{\odot}}{M}\right) \, 10^{6} \, \rm Hz$ \end{tabular} & \begin{tabular}{c} $\lesssim 2 \times 10^{-19} \, C^{5/2}  \, \left(\frac{\text{MHz}}{f}\right) \, \left(\frac{\text{Mpc}}{D}\right)$ \end{tabular} \\[1.4ex]
\cline{1-3}
\begin{tabular}{c} Superradiance: \\ annihilation \end{tabular} & \begin{tabular}{c} $\left( \frac{m_a}{10^{-9} \, \text{eV}} \right) 10^{6} \, \text{Hz} $ \end{tabular} & \begin{tabular}{c} $\lesssim 10^{-20}\left(\frac{\alpha}{l}\right)\epsilon \left(\frac{10 \, \mathrm{kPc}}{D} \right) \left(\frac{\text{MHz}}{f}\right)$ \end{tabular} \\[1.4ex]
\cline{1-3}
\begin{tabular}{c} Superradiance: \\ decay \end{tabular} & \begin{tabular}{c} $\left( \frac{m_a}{10^{-9} \, \text{eV}} \right) 10^{6} \, \text{Hz} $ \end{tabular} & \begin{tabular}{c} $\lesssim 3 \times 10^{-21} \epsilon^{1/2}\left(\frac{1 \, \text{MHz}}{f}\right)^{3/2} \, \left(\frac{10 \, \mathrm{kPc}}{D} \right)$ \end{tabular} \\[1.4ex]
\cline{1-3}
\hline
\end{tabular}}
\end{footnotesize}
\caption{Summary of coherent sources. The characteristic strain is given in Eqs.~\eqref{eq:hc} and~\eqref{eq:h0}. For all the details on how to obtain these expressions and the assumptions behind them, please check the corresponding sections: Sec.~\ref{sec:NS} for neutron star mergers, Sec.~\ref{sec:PBHmergers} for primordial BH mergers, Sec.~\ref{sec:ECOs} for exotic compact objects, Sec.~\ref{sec:Superradiance} for BH superradiance.}
\label{tab:summary-coherent}
\end{table}

\begin{table}[!ht]
\begin{footnotesize}
\centerline{
\begin{tabular}{|@{}c@{}||@{}c@{}|@{}c@{}|@{}c@{}|}
\hline
\begin{tabular}{c} Source \end{tabular} & \begin{tabular}{c} Frequency range \end{tabular} & \begin{tabular}{c} Amplitude \\ $\Omega_{\rm GW, 0}$ \end{tabular} & \begin{tabular}{c} Characteristic strain $h_c$ \\ (dimensionless) \end{tabular} \\[1.4ex]
\hline
\hline
\cline{1-4}
\begin{tabular}{c} Inflation: \\ vacuum amplitude \end{tabular} & \begin{tabular}{c} Flat in the range \\ $(10^{-16} - 10^{8}) \, {\rm Hz}$ \end{tabular} & \begin{tabular}{c} $\lesssim 10^{-16}$ \end{tabular} & \begin{tabular}{c} $\lesssim 10^{-32} \, \left(\frac{\text{MHz}}{f}\right)$ \end{tabular} \\[1.4ex]
\cline{1-4}
\begin{tabular}{c} Inflation: extra-species \end{tabular} & \begin{tabular}{c} $(10^5 - 10^{8}) \, {\rm Hz}$ \end{tabular} & \begin{tabular}{c} $\simeq 10^{-10}$ \end{tabular} & \begin{tabular}{c} $\lesssim 10^{-29} \, \left(\frac{\text{MHz}}{f}\right)$ \end{tabular} \\[1.4ex]
\cline{1-4}
\begin{tabular}{c} Inflation: \\ broken spatial reparametrization \end{tabular} & \begin{tabular}{c} Blue in the range \\ $(10^{-16} - 10^{8}) \, {\rm Hz}$ \end{tabular} & \begin{tabular}{c} $\simeq 10^{-10}$ \end{tabular} & \begin{tabular}{c} $\lesssim 10^{-29} \, \left(\frac{\text{MHz}}{f}\right)$ \end{tabular} \\[1.4ex]
\cline{1-4}
\begin{tabular}{c} Inflation: \\ secondary GW production \end{tabular} & \begin{tabular}{c} Flat or bump \end{tabular} & \begin{tabular}{c} $\lesssim 10^{-8}$ \end{tabular} & \begin{tabular}{c} $\lesssim 10^{-28} \, \left(\frac{\text{MHz}}{f}\right)$ \end{tabular} \\[1.4ex]
\cline{1-4}
\begin{tabular}{c} Preheating \end{tabular} & \begin{tabular}{c} $(10^{6}-10^9)$ {\rm Hz} \end{tabular} & \begin{tabular}{c} $\lesssim 10^{-10}$ \end{tabular} & \begin{tabular}{c} $\lesssim 10^{-29} \, \left(\frac{\text{MHz}}{f}\right)$ \end{tabular} \\[1.4ex]
\cline{1-4}
\begin{tabular}{c} Oscillons \end{tabular} & \begin{tabular}{c} $(10^{6}-10^9)$ {\rm Hz} \end{tabular} & \begin{tabular}{c} $\lesssim 10^{-10}$ \end{tabular} & \begin{tabular}{c} $\lesssim 10^{-29} \, \left(\frac{\text{MHz}}{f}\right)$ \end{tabular} \\[1.4ex]
\cline{1-4}
\begin{tabular}{c} Cosmic gravitational\\ microwave background \end{tabular} & \begin{tabular}{c} $f_{\rm peak} \sim (10-100)$ {\rm GHz} \end{tabular} & \begin{tabular}{c} $\Omega_{\rm GW}(f_{\rm peak})\lesssim 10^{-6}$ \end{tabular} & \begin{tabular}{c} $h_c(f_{\rm peak}) \lesssim 10^{-31} \, \left(\frac{\text{MHz}}{f}\right)$ \end{tabular} \\[1.4ex]
\cline{1-4}
\begin{tabular}{c} Phase transitions \end{tabular} & \begin{tabular}{c} $\lesssim 10^9 \,$ Hz \end{tabular} & \begin{tabular}{c} $\lesssim 10^{-8}$ \end{tabular} & \begin{tabular}{c} $\lesssim 10^{-28} \, \left(\frac{\text{MHz}}{f}\right)$ \end{tabular} \\[1.4ex]
\cline{1-4}
\begin{tabular}{c} Defects \end{tabular} & \begin{tabular}{c} Scale invariant \end{tabular} & \begin{tabular}{c} $\Omega_{\text{rad,0}} \, \frac{v^4}{M_p^4} \, F_U$ \end{tabular} & \begin{tabular}{c} $\times$ \end{tabular} \\[1.4ex]
\cline{1-4}
\begin{tabular}{c} Gauge textures \end{tabular} & \begin{tabular}{c} $\sim 10^{11} \, \frac{v}{M_p}  \, {\rm Hz}$ \end{tabular} & \begin{tabular}{c} $\sim 10^{-4} \frac{v^4}{{M_p}^4}$ \end{tabular} & \begin{tabular}{c} $\times$ \end{tabular} \\[1.4ex]
\cline{1-4}
\begin{tabular}{c} Grand unification \\ primordial BH evaporation \end{tabular} & \begin{tabular}{c} $(10^{18}-10^{15}) \,$ Hz \end{tabular} & \begin{tabular}{c} $\sim 10^{-8}$ \end{tabular} & \begin{tabular}{c} $\lesssim 10^{-28} \, \left(\frac{\text{MHz}}{f}\right)$ \end{tabular} \\[1.4ex]
\cline{1-4}
\hline
\end{tabular}}
\end{footnotesize}
\caption{Summary of stochastic sources. The expected amplitude is given in Eq.~\eqref{eq:OmegaAmplitude}, while the dimensionless characteristic strain is given in Eq.~\eqref{eq:charstrainstochastic}. The amplitudes reported are maximum values: for all the details on how to obtain these expressions, the dependence on the parameters of the models and the assumptions behind them, please check the corresponding sections: Sec.~\ref{sec:Inflation} for inflation, Sec.~\ref{sec:Preheating} for preheating and oscillons, Sec.~\ref{sec:CGMB} for the cosmic gravitational microwave background, Sec.~\ref{sec:EvaporatingPBHs} for primordial BH evaporation, Sec.~\ref{sec:PhaseTransitions} for phase transitions, Sec.~\ref{sec:TopologicalDefects} for topological defects and gauge textures, Sec.~\ref{sec:EvaporatingPBHs} for primordial BH evaporation.}
\label{tab:summary-stochastic}
\end{table}

\bibliographystyle{unsrt}
\bibliography{references}
\end{document}